\documentclass[11pt,letterpaper]{article}
\usepackage[dvipsnames,usenames,table]{xcolor}
\usepackage[colorlinks=true,urlcolor=blue,linkcolor=RoyalBlue,citecolor=OliveGreen]{hyperref}
\usepackage{amsmath,amsfonts,amssymb,amsthm,bbm} 
\usepackage[numbers,sort&compress]{natbib} 

\usepackage[margin=1in]{geometry}
\usepackage{graphicx}
\usepackage{transparent}
\usepackage{amsmath,amsfonts,amssymb}
\usepackage[algoruled,noend]{algorithm2e}
\usepackage{tikz}
\usepackage{times}
\usepackage{thm-restate}
\declaretheorem[name=Theorem]{theorem}
\declaretheorem[name=Lemma]{lemma}

\usepackage{cleveref}
\usepackage[normalem]{ulem}

\newcommand{\suppress}[1]{}

\newcommand{\nikhil}[1]{{\color{red} Nikhil: #1}}
\newcommand{\vasilis}[1]{{\color{blue} Vasilis: #1}}

\newcommand{\pp}{\mbox{\boldmath $p$}}

\newcommand{\seg}{\mbox{\rm seg}}
\newcommand{\segs}{\mbox{\rm segments}}
\newcommand{\valuex}{\mbox{\rm value}}
\newcommand{\rate}{\mbox{\rm rate}}

\newcommand{\good}{\mbox{\rm good}}

\newcommand{\ra}{\rightarrow}

\newcommand{\Q}{\mbox{\rm\bf Q}}
\newcommand{\Qplus}{\Q^+}

\newcommand{\CCP}{\textrm{\ref{CCP}}}

\newcommand{\R}{\mbox{\rm\bf R}}
\newcommand{\Rplus}{\R_+}

\makeatletter
\def\fnum@figure{{\bf Figure \thefigure}}
\def\fnum@table{{\bf Table \thetable}}

\long\def\@mycaption#1[#2]#3{\addcontentsline{\csname
 ext@#1\endcsname}{#1}{\protect\numberline{\csname
  the#1\endcsname}{\ignorespaces #2}}\par
     \begingroup
       \@parboxrestore
          \small
       \@makecaption{\csname fnum@#1\endcsname}{\ignorespaces
#3\endgroup}
      }

\newcommand{\SB}{c}

\newcommand{\UB}{d}

\newcommand{\lal}[1]{}
\newcommand{\aij}{ a_{ij}}
\newcommand{\xij}{ x_{ij}}
\newcommand{\uij}{ u_{ij}}
\newcommand{\vij}{ v_{ij}}
\newcommand{\bij}{ b_{ij}}
\newcommand{\cij}{ c_{ij}}

\newcommand{\grad}{\nabla}
\newcommand{\muij}{\mu_{ij}}

\newcommand{\budget}{B}
\newcommand{\srub}{\textsc{sr-ub}}

\begin{document}

\title{Convex Program Duality, Fisher Markets, \\ and Nash Social Welfare\thanks{Parts of this paper have earlier 
		appeared online as unpublished manuscripts under the titles: ``Fisher Markets 
and Convex Programs'' and ``New Convex Programs for Fisher's Market Model and its Generalizations''.}}


\author{
  Richard Cole \thanks{Courant Institute, New York University, New York City. cole@cs.nyu.edu }
  \and
	Nikhil R. Devanur \thanks{Microsoft Research, Redmond. nikdev@microsoft.com }
	\and
  Vasilis Gkatzelis \thanks{College of Computing \& Informatics, Drexel University, Philadelphia. gkatz@drexel.edu }
	\and
	Kamal Jain \thanks{Faira, Seattle. kamaljain@gmail.com }
	\and
	Tung Mai \thanks{College of Computing, Georgia Institute of Technology, Atlanta. maitung89@gatech.edu}
	\and 
	Vijay V. Vazirani \thanks{College of Computing, Georgia Institute of Technology, Atlanta. vazirani@cc.gatech.edu }
	\and
	Sadra Yazdanbod \thanks{College of Computing, Georgia Institute of Technology, Atlanta. yazdanbod@gatech.edu}
}

\date{}
\maketitle
\thispagestyle{empty}

\begin{abstract}
	We study Fisher markets and the problem of maximizing the Nash social welfare (NSW), and show several closely related new results. In particular, we obtain:
	\begin{itemize}
		\item  A new integer program for the NSW maximization problem 
		whose fractional relaxation has a 
	 bounded integrality gap. In contrast, the natural integer program has an unbounded integrality gap.
		\item  An improved, and tight, factor 2 analysis of the algorithm of \cite{CG15}; in turn showing that the integrality gap of the above relaxation is at most 2.  The  approximation factor shown by \cite{CG15} was $2e^{1/e}\approx 2.89$. 
		\item A lower bound of $e^{1/e}\approx 1.44$ on the integrality gap of this relaxation. 
		\item New convex programs for natural generalizations of linear Fisher markets and proofs that these markets admit rational equilibria. 
	\end{itemize}
	
	These results were obtained by establishing connections between previously known disparate results, and they help uncover their mathematical underpinnings. 
We show a formal connection between the convex programs of Eisenberg and Gale and that of Shmyrev, namely that their duals are equivalent up to a change 
of variables. Both programs capture equilibria of linear Fisher markets. By adding suitable constraints to Shmyrev's program, we obtain a convex program 
that captures equilibria of the spending-restricted market model defined by \cite{CG15} in the context of the NSW maximization problem. Further, adding 
certain integral constraints to this program we get the integer program for the NSW mentioned above. 
	
	The basic tool we use is convex programming duality. In the special case of convex programs with linear constraints (but convex objectives), we show a particularly simple way of obtaining dual programs, putting it almost at par with linear program duality. This simple way of finding duals has been used subsequently for many other applications. 
%
\end{abstract}

\newpage

\setcounter{page}{1}

\suppress{
\textcolor{red}{Template:\\ 
State of the art: NSW maximization, spending restricted eq., upper bound, factor 2.8. \\
Questions: Is there a convex program for the s.r. eq? \\
Is there an integer program with a bounded integrality gap for NSW max? \\
Factor improved? for this algo? other algos? \\
Another thread: EG and Shmyrev CPs, both find equilibria of linear Fisher. Is there a connection?\\
Convex programs for other Fisher markets such as quasi-linear utilities, spending constraint utilities, transaction costs, linear AD? \\
Contributions: \\
simple duality theory. Applications beyond this paper (currenty 4th para in the intro). \\
connections between EG, Shmyrev. \\
CP for spending restricted markets. \\
Equivalence of upper bound on NSW and objective of CP. \\
Integral constraints capture NSW. \\
Tight analysis of 2. \\
Lower bound of 1.44. \\
Introduce Utility restricted markets and CP for it. Also independently by 	\cite{bei2016computing}\\
Other generalizations: QL. Reserve prices.  In appendix?\\
Open Questions. 
}
\newpage
}

\section{Introduction}
\label{sec.intro}
Recently, \citet{CG15} gave the first constant factor approximation algorithm for the problem of maximizing the Nash social welfare 
(NSW). In this problem, a set of indivisible goods needs to be allocated to agents with additive utilities, and the goal is to
compute an allocation that maximizes the geometric mean of the agents' utilities. The natural integer program for this problem
is closely related to the Fisher market model: if we relax the integrality constraint of the allocation, i.e., assume that the
the goods are divisible, this program reduces to the Eisenberg-Gale (EG) convex program \cite{eisenberg}, whose solutions correspond
to market equilibria for the linear Fisher market. Therefore, a canonical approach for designing a NSW approximation algorithm would 
be to compute a fractional allocation via the EG program, and then ``round'' it to get an integral one. However, \cite{CG15} observed
that this program's integrality gap is unbounded, and they were forced to follow an unconventional approach in analyzing
their algorithm. This algorithm used an alternative fractional allocation, the {\em spending-restricted} (SR) equilibrium, 
and they had to come up with an independent upper bound of the optimal NSW in order to prove that the approximation factor is 
at most $2e^{1/e}\approx 2.89$.

The absence of a conventional analysis for this problem could be, in part, to blame for the lack of progress on important 
follow-up problems (e.g., see Section~\ref{sec:Discussion}). For instance, the SR equilibrium introduces constraints that 
are incompatible with the EG program, so~\cite{CG15} had to use a complicated algorithm for computing this allocation. 
Generalizing such an algorithm may be non-trivial, and so would proving new upper bounds for the optimal NSW.
In this paper we remove this obstacle by uncovering the underlying structure of the NSW problem and shedding
new light on the results of~\cite{CG15}. Specifically, we propose a new integer program which, as we show, also computes 
the optimal NSW allocation. More importantly, we prove that the relaxation of this program computes the SR equilibrium, 
and, quite surprisingly, we also show that the objective of this program happens to be precisely the upper bound that was 
used in~\cite{CG15}. As a result, this new integer program yields a convex program for computing the SR equilibrium and,
unlike the standard program, it has an integrality gap that is bounded by $2.89$. In addition to this, we give a family 
of instances showing a lower bound of $e^{1/e}\approx 1.44$ on the integrality gap, and we provide a tight analysis of the 
algorithm of~\cite{CG15} to show that its approximation factor is $2$, which also puts an upper bound of $2$ on the 
integrality gap of the new program. 

Apart from the results regarding the NSW problem, we also reveal interesting connections between seemingly
disparate results, and we provide convex programs for computing market equilibria in interesting generalizations 
of Fisher's market model. For instance, besides the EG program, there is another very different convex program for the 
linear Fisher market, due to Shmyrev \cite{shmyrev2009algorithm}; however, there were no known connections between these 
two programs. Using our techniques, we show that one can define a dual program for each of them, and the 
two duals are the same, up to a change of variables. Furthermore, by adding suitable constraints to Shmyrev's program, 
we obtain the convex program that captures the SR equilibria.

The spending-restricted market model is a generalization of Fisher's market model and has potential use beyond its
NSW application. Under this model, sellers can declare an upper bound on the money they wish to earn in the market 
(and take back their unsold good). Therefore, the total amount of money that the buyers can spend on this seller's
good is bounded. Assume that each seller is selling his services in the market. In the last half 
century, society has seen the emergence of a multitude of very high end jobs which call for a lot of expertise and in turn pay very large salaries.
Indeed, the holders of such jobs do not need to work full time to make a comfortable living and one sees  numerous such people preferring to work for
shorter hours and having a lot more time for leisure. High end dentists, doctors and investors fall in this category. 
The spending restricted model allows such agents to specify a limit on their earnings beyond which they do not wish
to sell their services anymore. 
    
Another generalization of the linear Fisher model that we study is the {\em utility restricted} (UR) model. 
In this model, buyers can declare an upper bound on the amount of utility they wish to derive (and take 
back the unused part of their money). This model is natural as well: in thrift, it is reasonable to assume that a 
buyer would only want to buy goods that are absolutely necessary, i.e., place an upper bound amount on 
utility, and not spend all of her money right away.

Thus, in the SR model, the supply of a good is a function of the prices and, in the UR model,
the amount of money a buyer spends in the market is a function of the prices.
In the presence of these additional constraints, do equilibria exist and can they be computed in polynomial time? 
We give a convex program for the second model as well, this time by generalizing the EG program.
Existence of equilibria for both models follows from these convex programs. 
We further show that both models admit rational equilibria, i.e., prices and allocations are rational numbers 
if all parameters specified in the instance are rational. As a consequence, the ellipsoid algorithm will 
find a solution to the convex programs in polynomial time. 

For some of the results listed above, the techniques that we use are based on convex program duality.
We consider a special class of convex programs, those with convex objective functions and {\em linear} constraints,
and show that the duals can be constructed using a simple set of rules,\footnote{The dual is obtained using the usual Lagrangian 
relaxation technique. We show a ``short-cut" for applying this technique, making it especially easy to derive the 
dual for the special case we consider.} which are almost as simple as those for linear programs. 
We note that convex programming duality  is usually stated in its most general form, with convex objective functions and 
convex constraints, e.g., see the excellent references by  \citet{BV} and \citet{rockafellar1970convex}.
At this level of generality the process of constructing the dual of a convex program is quite tedious.
Following an earlier version of this paper\footnote{The part of the current paper about convex programming duality 
had been made available online since 2010 as the following unpublished manuscript: N. R. Devanur, Fisher Markets and 
Convex Programs. The manuscript is now incorporated into this paper.}, these rules have found serveral additional 
applications in deriving convex programs: for Fisher markets under spending constraint utilities~\cite{BDXnikhil}, 
Fisher markets with transaction costs~\cite{CDK10}, Arrow-Debreu market with linear utilities \cite{devanur2013rational}, 
and Fisher markets with reserve prices \cite{cole2015does}.
They have also been used in the design of algorithms: for simplex-like algorithms for spending constraint utilities 
and perfect price discrimination markets \cite{garg2013towards}, in analyzing the convergence of the tatonnement 
process~\cite{cheung2013tatonnement}, in designing online algorithms for scheduling~\cite{im2014selfishmigrate,buchbinder2014online,devanur2014primal}, 
and online algorithms for welfare maximization with production costs~\cite{huang2015welfare}. Finally, they have 
also been used in bounding the price of anarchy of certain games \cite{kulkarni2015robust}.

\suppress{
We will say that $\pp$
are {\em market clearing prices} if after each $i$ is given an optimal bundle, there is
no deficiency or surplus of any good, i.e., the market clears.

We will call each step of $f_j^i$ a {\em segment}. The set of segments defined in function
$f_j^i$ will be denoted $\seg(f_j^i)$.
Suppose one of these segments, $s$, has
range $[a, b] \subseteq [0, e(i)]$, and $f_j^i(x) = c$, for $x \in [a, b]$. Then, we will
define $\valuex(s) = b-a$, $\rate(s) = c$, and $\good(s) = j$; we will assume
that good 0 represents money. Let
$\segs(i)$ denote the set of all segments of buyer $i$, i.e.,
\[ \segs(i) = \bigcup_{j=0}^n  {\seg(f_j^i)}  .\]

Let us assume that the given problem instance satisfies the following (mild) conditions:
\begin{itemize}
\item
For each good, there is a potential buyer, i.e.,
\[ \forall j \in A \ \exists i \in B \ \exists s \in \seg(f_j^i) \ : \  \rate(s) > 0 .\]

\item
Each buyer has a desire to use all her money (to buy goods or to keep some unspent), i.e.,
\[ \forall i \in B :   \sum_{s \in \segs(i), \ \rate(s) > 0}  {\valuex(s) \geq e(i)} .\]
\end{itemize}
}

\section{Preliminaries}
\label{sec.model}
\suppress{
\textcolor{red}{Template:\\ 
Fisher market. linear utilities. QL utils. \\
SR market. UR market. 	\\
NSW problem, SR market with $q_i = 1$. upper bound.  \\ 
}
}

Fisher's market model is the following: let $M$ be a set of $m$ divisible goods and $N$ be a set of $n$ buyers.
Each buyer $i$ comes to the market with a budget of $\budget_i$
and we may assume
w.l.o.g.\ that the market has one unit of each good.
Each buyer $i$ has a utility function,
$u_i: \Rplus^m \ra \Rplus$, giving the utility that $i$ derives from each bundle of goods.
The utility of buyer $i$ is said to be {\em linear} if
there are parameters $v_{ij} \in \Rplus$, specifying the value derived by $i$ from one unit of good $j$.
Her utility for the entire bundle is additive, i.e., $u_i(x) = \sum_{j \in M} {v_{ij} x_{ij}}$.
Utility function $u_i$ is said to be {\em quasi-linear} if, agents have utility for the money spent as well, i.e., 
$u_i(x)= \sum_{j \in M} {(v_{ij}-p_{j}) x_{ij}}$.
Utility function $u_i$ is said to be {\em Leontief} if,
given parameters $a_{ij} \in \Rplus \cup \{0\}$ for each good $j \in M$,  
 $u_i(x) = \min_{j \in M} {x_{ij}/a_{ij}}$.
Finally, $u_i$ is said to be {\em constant elasticity of substitution (CES) with parameter $\rho$} if 
given parameters $\alpha_j$ for each good $j \in M$,  
$u_i(x) = \left( \sum_{j=1}^m \alpha_j x_j^{\rho} \right)^{{1 \over {\rho}}}$.
Throughout the main body of the paper we assume that the utilities are linear unless we note otherwise.

\paragraph{Market equilibrium:} 
Let $p_j\in \Rplus$  be the price of good $j$ and $\xij\in \Rplus$ denote the amount of good $j$ allocated to buyer $i$. 
(We use $p$ and $x$ to denote the vectors of all prices and allocations, respectively.)
These are said to form an {\em equilibrium} if the following conditions hold.
\begin{enumerate}
	\item The allocation of each buyer $i$ maximizes his utility, subject to her budget constraint, $\sum_j p_j \xij \leq \budget_i. $
	\item Each good $j$ that has a price $p_j >0$ is allocated fully, i.e., $\sum_i \xij = 1$. A good is allowed to have price $p_j = 0$ as long as $\sum_i \xij \leq 1$. 
\end{enumerate} 
%

Two natural generalizations of Fisher's model that we consider are the following.
In the first model which we call {\em Spending-Restricted} (SR) model,
each seller $j$ has an upper bound $\SB_j$ on the amount of money $j$ wants to earn in
the market. Once he earns $\SB_j$, selling the least amount of his good, he wants to take back the unsold
portion of his good. In other words, the amount of money spent on the good of seller $j$ is restricted by $\SB_j$.
In equilibrium, buyers spend all their money and get an optimal bundle of goods.
Formally, the second equilibrium condition above is modified to $\forall j \in M,   \sum_i \xij \leq 1$, and $\sum_i p_j \xij \leq c_j$, and either
\[\textstyle \sum_i \xij = 1, \text{ or } \sum_i p_j \xij = c_j, \text{ or } p_j = 0. \]
In the second model which we call {\em Utility-Restricted} (UR) model,
buyers have upper bounds $\UB_i$ on the utility they want to derive in the market. Once buyer $i$ derives utility 
$\UB_i$, spending the least amount of money at prices $p$, she wants to keep the left-over money.
In other words, the utility of buyer $i$ is restricted by $\UB_i$.
In equilibrium, each good with a positive price should be fully sold.
Formally, the first equilibrium condition is modified to $\forall i \in N, u_i(x) \leq d_i,$ and  $\sum_j p_j \xij \leq B_i$, and either 
\[\textstyle  x \text{ minimizes } \sum_j p_j \xij \text{ s.t. } u_i(x) = d_i, \text{or maximizes } u_i(x) \text{ s.t. } \sum_j p_j \xij \leq \budget_i. \]

Given an equilibrium $(p,x)$, we denote the total money spent on item $j$ by $q_j$,
and the money that agent $i$ spends on item $j$ by $b_{ij}$.
The {\em spending graph}, $Q(b)$, of a given spending vector $b$, is a bipartite graph 
where the set of agents corresponds to vertices of one side of the graph and the set of 
items corresponds to vertices of the other side. Each agent $i$ is connected to the 
items that she spends money on, i.e., there is an edge between $i$ and $j$ if and only 
if $b_{ij}>0$. Note that each agent only spends money on the set of her maximum ``bang 
per buck" items, i.e., the set of items that maximize $v_{ij}/p_j$. Therefore, by assuming 
some unique tie breaking rule among goods we can rearrange the spending to ensure that the 
spending graph is a forest of trees. Throughout this paper we assume that the spending graph 
is always a forest of trees.\vspace{ -2 mm}
\paragraph{Nash Social Welfare:}
Given a  set $M$ of $m$ indivisible items and a set $N$ of $n$ agents, 
an \emph{integral} allocation of items to agents restricts the allocation $\xij$ to lie in the set $\{0,1\}$.
The {\em Nash social welfare} (NSW) (also known as Bernoulli-Nash social welfare)  of 
an {integral} allocation $x$ is defined as the geometric mean of the agents utilities, i.e., 
$(\prod_{i\in N} u_i(x))^{1/n}$ \cite{KN79,nash.bargain}.
The NSW maximization problem is to find an integral allocation that maximizes the NSW. 
(We may assume w.l.o.g. that $n\leq m$ for this problem.)
\citet{CG15} considered this problem when agents have linear utilities, and gave a 
$2e^{1/e}\approx 2.89$ factor approximation for it.
We now state the upper bound on the optimum value that is used in their result. 

Consider an SR market with the same items and agents and utilities.
Suppose the items are divisible and have spending restriction of 1 on all items,
i.e., $\forall j\in M$, $\SB_j=1$. Let $\bar{x}$ and $\bar{p}$ be an equilibrium allocation and 
price vector of the market.
Note that multiplying all the $\vij$ values of a given agent $i$ by the same positive number does not
change the optimal solution or the approximation factor for the problem. 
In an equilibrium allocation all goods allocated to an agent must have the same ``bang per buck'' ratio $\vij/\bar{p}_j$ (as was shown in \cite{CG15}).
We can therefore normalize each agent's valuations so that $v_{ij} =\bar{p}_j$ if $\bar{x}_{ij}>0$, without loss of generality. 
We henceforth assume that the valuations are normalized this way in every NSW problem instance.
Given such a scaling, we define the following quantity which was used in~\cite{CG15} as an upper bound on 
the optimal NSW value.
\[\textstyle \srub := \left(\prod_{j \in M : \bar{p}_j \geq 1} \bar{p}_j \right)^{1/n} .\]
We now state the following lemma that is proved by \cite{CG15}. 
\begin{lemma}[\cite{CG15}]\label{upperbound}
For linear utilities, $\textstyle \max_{\xij\in\{0,1\}}\left( \prod_{i\in N} u_i(x)\right)^{1/n} \leq \srub. $
\end{lemma}

\section{Convex programming duality}\label{sec:CPduality}
\subsection{Fenchel Conjugate}\vspace{-2 mm}
We now define the {\em Fenchel conjugate} of a function,  and note some of its properties; 
see \citet{rockafellar1970convex} for a detailed treatment.
This will be the key ingredient in extending the simple set of rules for LP duality to convex programs.
Suppose that $f: \mathbb{R}^n \rightarrow \mathbb{R}$ is a function.
The conjugate of $f$ is  $f^*:\mathbb{R}^n \rightarrow\mathbb{R} $ and is defined as
 $ f^*(\mu) := \sup_x\{ \mu^Tx - f(x) \}$.
Although the conjugate is defined for any function $f$, for the rest of the article we will assume
that $f$ is {\em strictly convex and differentiable}, since this is the case that is most interesting to the applications we discuss.

\noindent {\bf Properties of $f^*$:} We note some useful properties here. See Appendix \ref{app:convexduality} for more properties. 
\begin{itemize}
  \item[$\bullet$] If $\mu$ and $x$ are such that $f(x) + f^*(\mu) = \mu^Tx$ then
$\grad f(x) = \mu $ and $\grad f^*(\mu) = x$.
\item[$\bullet$] Vice versa, if $\grad f(x) = \mu $ then $\grad f^*(\mu) = x$ and
$f(x) + f^*(\mu) = \mu^Tx$.
\end{itemize}
We say that $(x,\mu)$ form a complementary pair w.r.t.\ $f$ if they satisfy either one of these two conditions.

\subsection{Convex programs with linear constraints}
Suppose that we have a convex program with a convex/concave objective function and linear constraints. 
We can derive another convex program that is the {\em dual} of this, using Lagrangian duality.
This is usually a long calculation. The goal of this section is to identify a shortcut for the same.
\begin{restatable}{lemma}{CPduality}\label{lem:CPduality}
	The following pairs of convex programs are duals of each other, i.e., the optimum of the primal is at most the optimum of the dual  (weak duality). 
If the primal is infeasible, then the dual is unbounded (and vice versa). 
	
\begin{minipage}{0.35\textwidth}  
	\[\text{Primal: }\max \textstyle \sum_i c_i x_i - f(x) \text{ s.t. }\]
	\[ \forall~ j,\textstyle  \sum_i \aij x_i \leq b_j,\]
\end{minipage}
\hfill\vline\hfill
\begin{minipage}{0.35\textwidth}  
	\[\text{Dual: } \min \textstyle \sum_j b_j\lambda_j + f^*(\mu) \text{ s.t.} \]
	\[\forall~ i,\textstyle  \sum_j \aij \lambda_j = c_i - \mu_i,\]
	\[\forall~ j, \lambda_j \geq 0 .\]
\end{minipage}

\noindent If the primal constraints are strictly feasible, i.e., 
there exists $\hat{x}$ such that for all $j$  $\sum_i \aij \hat{x}_i < b_j$,
then the two optima are the same (strong duality) and the following  generalized complementary slackness conditions characterize them:
 \begin{itemize}
 	\item $x_i > 0 \Rightarrow \sum_j\aij \lambda_j= c_i - \mu_i$, $\quad \quad$ $\lambda_j > 0 \Rightarrow \sum_i \aij x_i = \budget_i$ and
 	\item $x$ and $\mu$ form a complementary pair wrt $f$, i.e.,
 	$\mu = \grad f(x), x = \grad f^*(\mu)$ and $f(x) + f^*(\mu) = \mu^Tx$.
 \end{itemize}
\end{restatable}
The proofs of all  lemmas in this section are in Appendix \ref{app:convexduality}. Note the similarity to LP duality. 
When an LP is infeasible the dual becomes unbounded. The same happens with these convex programs as well. 
The differences are as follows.
Suppose the concave part of the primal objective is  $-f(x)$. There is an extra variable $\mu_i$ for every
variable $x_i$ that occurs in $f$. In the constraint corresponding to $x_i$, the term $-\mu_i$ appears on the RHS along with the constant term.
Finally the dual objective has $f^*(\mu)$ in addition to the linear terms. In other words, we {\em relax}  the constraint
corresponding to $x_i$ by allowing a slack of $\mu_i$, and {\em charge} $f^*(\mu)$ to the objective function.

Similarly, the primal program with non-negativity constraints on variables and the corresponding dual program take the following form.

\begin{minipage}{0.35\textwidth}  
	\[\textstyle \text{Primal: }\max \textstyle \sum_i c_i x_i - f(x) \text{ s.t. }\]
\[ \textstyle \forall~ j,\textstyle  \sum_i \aij x_i \leq b_j,\]
\[ \textstyle  \forall~ i, x_i \geq 0 .\]
\end{minipage}
\hfill\vline\hfill
\begin{minipage}{0.35\textwidth}  
\[\textstyle \text{Dual: } \min \textstyle \sum_j b_j\lambda_j + f^*(\mu) \text{ s.t.} \]
\[\textstyle \forall~ i,\textstyle  \sum_j \aij \lambda_j \geq c_i - \mu_i,\]
\[\textstyle \forall~ j, \lambda_j \geq 0 .\]
\end{minipage}

\noindent The dual of a minimization program has the following form. 

\begin{minipage}{0.35\textwidth}  
	\[\textstyle \text{Primal: }\min \sum_i c_i x_i + f(x) \text{ s.t. }\]
\[ \textstyle \forall~ j, \sum_i \aij x_i \geq b_j,\]
\[ \textstyle \forall~ i, x_i \geq 0 .\]
\end{minipage}
\hfill\vline\hfill
\begin{minipage}{0.35\textwidth}  
\[\textstyle \text{Dual: }\max \sum_j b_j\lambda_j - f^*(\mu) \text{ s.t.} \]
\[\textstyle \forall~ i, \sum_j \aij \lambda_j \leq c_i + \mu_i,\]
\[\textstyle \forall~ j, \lambda_j \geq 0 .\]
\end{minipage}

\section{Convex programs for Fisher markets}
\label{sec.4}
%
We now use the technology developed in the previous section to show a formal connection between the Eisenberg-Gale and 
Shmyrev convex programs, both of which are known to capture equilibria of linear Fisher markets as their optima. 
As a first step we construct the dual of the Eisenberg-Gale convex program.
\begin{restatable}{lemma}{EGdual}\label{lem:EGdual}
	The following pairs of convex programs are duals of each other. The dual variables $p_j$ of 
	an optimal solution are equilibrium prices of the corresponding linear Fisher market.  
		
	\begin{minipage}{0.35\textwidth}  
	\[ \textstyle  \text{EG Program: } \max \sum_{i} \budget_i \log u_i \text{ s.t.} \]
	\[ \textstyle\forall~ i, u_i \leq \sum_j \vij \xij ,\]
	\[\textstyle \forall~j, \sum_i \xij \leq 1, \]
	\[ \xij \geq 0.\]
	\end{minipage}
	\hfill\vline\hfill
	\begin{minipage}{0.35\textwidth}  
\begin{equation}\label{cp.dualeg} \textstyle\min \sum_j p_j  - \sum_i \budget_i \log (\beta_i) \text{ s.t.}\end{equation}
\[ \forall~i,j, p_j \geq \vij\beta_i.\]
	\[ \enspace\]
		\[ \enspace\]
	\end{minipage}
\end{restatable}

In fact, we can even eliminate the $\beta_i$'s by observing that in an optimal solution,
$\beta_i = \min_j \left\{ p_j/\vij \right \}$. This gives a convex (but not strictly convex)
function of the $p_j$'s that is minimized at equilibrium. Note that this is an unconstrained\footnote{Although with some analysis, one can derive that the optimum solution satisfies that $p_j \geq 0$, and 
$\sum_j p_j = \sum_i \budget_i$, the program itself has no constraints.} minimization. 
The function is $ \sum_j p_j  - \sum_i \budget_i \log (\min_j \left\{ p_j/\vij \right \}).$
An interesting property of this function is that the (sub)gradient of this function at any price vector corresponds to 
the (set of) excess supply of the market with the given price vector. This implies that a tattonement style price update,
where the price is increased if the excess supply is negative and is decreased if it is positive, is actually 
equivalent to gradient descent.  This fact was used to analyze the convergence of the tatonnement process in \cite{cheung2013tatonnement}. 
A convex program that is very similar to (\ref{cp.dualeg}) was also discovered 
independently by Garg \cite{jgarg08}. However it is not clear how they arrived at it, or 
if they realize that this is the dual of the Eisenberg-Gale convex program.
Going back to Convex Program (\ref{cp.dualeg}), we  write an equivalent program by taking the $\log$s in each of the constraints.
\[\textstyle \min \sum_j p_j  - \sum_i \budget_i \log (\beta_i) \text{ s.t}\]
\[\textstyle \forall~i,j, \log p_j \geq \log \vij + \log \beta_i.\]
Replacing $q_j = \log p_j $ and $\gamma_i = -\log \beta_i $ as the variables, we get the following convex program (\ref{cp.dualeglogs}), and its dual  ($\CCP$).

\begin{restatable}{lemma}{Shmyrev}\label{lem:Shmyrev}
	The following convex programs are duals of each other. 
	
		\begin{minipage}{0.35\textwidth}  
		\begin{equation}\label{cp.dualeglogs}\textstyle \min \sum_j e^{q_j}  + \sum_i \budget_i \gamma_i \text{ s.t.}\end{equation}
		\[ \forall~i,j, \gamma_i + q_j  \geq \log \vij.\]
		\[ \enspace \]
		\[ \enspace \]		
		\end{minipage}
		\hfill\vline\hfill
		\begin{minipage}{0.55\textwidth}  
		\[ \textstyle\max \sum_{i,j} \bij \log \vij - \sum_j (p_j \log p_j - p_j) \text{ s.t.} \tag{CP}\label{CCP}\]
		\[ \textstyle\forall~j, \sum_i \bij = p_j, \]
		\[ \textstyle\forall~ i, \sum_j \bij = \budget_i,\]
		\[\textstyle \forall~ i,j,\bij \geq 0. \]
		\end{minipage}

\end{restatable}
 By abuse of notation, we use $p_j$ for the variables in ($\CCP$) since it turns out that these once again correspond to equilibrium prices. 
We can remove the $-p_j$ at the end of the objective in ($\CCP$) since the constraints imply that $\sum_j p_j = \sum_i \budget_i$, which is a constant.
On removing these terms, we get the convex program of Shmyrev \cite{shmyrev2009algorithm}.  Thus (\CCP) and EG convex programs 
have the same dual, modulo a change of variables! 
%

\paragraph{Quasi-linear utilities:}
For some markets, it is not clear how to generalize the Eisenberg-Gale convex program, 
but the dual generalizes easily, and the optimality conditions can be easily seen to be 
equivalent to equilibrium conditions. We now show an example of this. 
Recall that a buyer $i$  has  a quasi-linear utility if it is of the form $\sum_j (\vij - p_j)\xij$.
In particular, if all the prices are such that $p_j > \vij$, then the buyer prefers to not be allocated any good and go
back with his budget unspent. It is easy to see that the following convex program (\ref{cp.dualegQuasilinear}) captures  equilibrium prices for such utilities.
In fact, given this convex program, one could take its dual to get an EG-type convex program as well. Although this is a small modification of the EG program, it is not clear how one would arrive at this directly without going through the dual. 
\begin{lemma}\label{lem:QL}
	The following pairs of convex programs are duals of each other, and capture the equilibria of Fisher markets with quasi-linear utilities as their optima. 
	
	\begin{minipage}{0.35\textwidth}  
	\begin{equation}\label{cp.dualegQuasilinear} \textstyle\min \sum_j p_j  - \sum_i B_i \log (\beta_i) \text{ s.t.}\end{equation}
	\[ \textstyle\forall~i,j, p_j \geq \vij\beta_i,\]
	\[\textstyle \forall~i, \beta_i \leq 1. \]
		\[ \enspace \]		
	\end{minipage}
	\hfill\vline\hfill
	\begin{minipage}{0.35\textwidth}  
	\[ \textstyle\max \sum_{i} B_i \log u_i -v_i \text{ s.t.} \]
	\[ \textstyle\forall~ i, u_i \leq \sum_j \vij \xij + v_i ,\]
	\[ \textstyle\forall~j, \sum_i \xij \leq 1, \]
	\[\textstyle \forall i,j,~~\xij ,v_i \geq 0 .\]
	\end{minipage}
\end{lemma}

\paragraph{Summary and Extensions:} 
In this section we showed two applications of the convex programming duality in \Cref{sec:CPduality}, the relation between the EG and Shmyrev convex programs, and a convex program for Quasi-linear utilities. We mention other applications of this tool in the introduction, some of which are in \Cref{app:convexduality}. 
We give a convex program that captures SR equilibrium, and study existence, uniqueness and rationality of equilibrium in \Cref{existence}. 
Further, the same analysis can be extended to what are called spending constraint utilities (\Cref{sec:m2sc}). 
We do the same (convex programs, existence, uniqueness and rationality) for UR markets with linear, Leontief and CES utilities in \Cref{utility_bound_section}. 
The convex program for the SR model is closely related to NSW maximization, as we will discuss in the next section.
\vspace{-2 mm}

\section{A new program for the Nash social welfare problem}\label{sec.NSW}\vspace{-2 mm}
In this section we focus on the APX-hard problem of maximizing the NSW with indivisible items~\cite{CG15,Lee15}. 
When the agents have 
linear valuations, this problem has a natural representation as a convex program (see program on the left 
below). In this program, there is a variable $x_{ij}$ for each agent $i$ and item $j$ and its value is 
either 0 or 1, depending on whether the agent is allocated the item or not. An appealing property of this 
program is that, if we relax the constraint that $x_{ij}\in\{0, 1\}$, then the program reduces to the 
Eisenberg-Gale program\footnote{To verify this fact, apply a logarithmic transformation to the objective.}, 
which can be solved in polynomial time. This opens the way for a standard approach for designing an 
approximation algorithm: compute the fractional allocation using the EG program and then use a rounding 
algorithm to get a good integral allocation. Unfortunately, as was shown in~\cite{CG15}, the integrality 
gap of this program is unbounded, so this approach is doomed to fail. 

Facing the unbounded integrality gap obstacle, \cite{CG15} take a non-standard approach in designing 
an approximation algorithm. Motivated by the market equilibrium interpretation of the EG program, they 
propose the spending-restricted equilibrium, and they then independently prove an upper bound for the 
optimal NSW value (which we call $\srub$, see Lemma \ref{upperbound}). They then ``round'' the fractional 
allocation implied by the SR equilibrium, and compare the NSW of the rounded solution to \srub. In this section, we propose 
a new integer program, which we refer to as the {\em spending-restricted} (\ref{SR}) program (see program 
on the right below)\footnote{The \ref{SR} program is not, strictly speaking, presented as an integer program, 
but we could introduce a new variable $a_j$ for each item $j$ and replace the constraint $b_{ij}\in \{0, q_j\}$ 
with the constraints $b_{ij}=a_{ij} q_j$ and $a_{ij}\in\{0, 1\}$ to make it an integer program.}, and show the following results.
\begin{itemize}
	\item The optimal solution of the \ref{SR} program corresponds to the NSW maximizing integral allocation, and the optimal objective function value of this program is equal to the optimal NSW value. 
	\item The fractional relaxation of this program computes the SR equilibrium. 
	\item The objective value of the fractional relaxation is equal to the upper bound $\srub$. 
	\item This relaxation therefore has an integrality gap of at most $2e^{1/e}\approx 2.89$. We also show a lower bound of $e^{1/e}\approx 1.44$ on this integrality gap.  
\end{itemize}

\begin{minipage}{0.35\textwidth}  
\[ \textstyle  \text{max } \left(\prod_i u_i\right)^{1/n} \text{ s.t.} \]
\[ \textstyle \forall i,~ ~ u_i = \sum_j x_{ij} v_{ij} \]
\[ \textstyle \forall j, ~~ \sum_i x_{ij} =1 \]
\[ \textstyle \forall i, j, ~~ x_{ij}\in \{0, 1\}. \]
\end{minipage}
\hfill\vline\hfill
\begin{minipage}{0.40\textwidth}  
\[ \textstyle \text{max } ~~ \left(\frac{\prod_i\prod_j v_{ij}^{b_{ij}}}{\prod_j q_j^{q_j}}\right)^{1/n}  \text{ s.t.} \tag{SR} \label{SR}\]
\[ \textstyle \forall j, ~~ \sum_i b_{ij} =q_j \]
\[ \textstyle \forall i, ~~ \sum_j b_{ij} =1 \]
\[ \textstyle  \forall i, j, ~~ q_j \leq 1, b_{ij}\in \{0, q_j\} \]
\end{minipage}


Unlike the standard program for the NSW problem, the \ref{SR} program uses variables $q_j$ and 
$b_{ij}\in \{0, q_j\}$. Any solution to this program, corresponds to an allocation of 
indivisible items to agents. In particular, an agent $i$ is allocated an item 
$j$ if and only if $b_{ij}=q_j$.\footnote{Note that we can assume $\forall j, ~q_j>0$ in an 
equilibrium w.l.o.g. because if $q_j=0$ then the equilibrium conditions imply the value of item 
$j$ is zero for all agents.} If we relax the constraint that $b_{ij}\in \{0, q_j\}$ and 
apply a logarithmic transformation of the objective function, we get a convex program, which 
we can compute in polynomial time. We call this relaxation the f-SR program. Note that the 
spending constraint ($q_j\leq 1$) is not binding in the SR program, but this is not true for f-SR.

The following lemma shows that the two programs above do, in fact, compute the same allocation.

\begin{lemma}\label{lem:SR_OPT}
The optimal solution of the \ref{SR} program corresponds to the NSW maximizing allocation
of indivisible items to agents. The objective function value of this solution is equal to
the optimal NSW value. 
\end{lemma}
\begin{proof}
Suppose that we fix the integral choices, i.e., for each $i$ and $j$ we fix whether $b_{ij} = 0$ or $b_{ij} = q_j$. 
For all $j$,  due to the constraint that $\sum_{i} b_{ij} = q_j$, there can only be one $i$ such that  $b_{ij} = q_j$.  Hence determining the integral choices is equivalent to determining an integral allocation. 
Let $S_i$ denote the set of items allocated to $i$ in this integral allocation.   
We show that given these integral choices,  
setting $b_{ij}=\frac{v_{ij}}{\sum_{k \in S_i} v_{ik}}$
makes the objective function equal to the NSW of the allocation, and this is indeed the optimal (objective maximizing) choice of these variables. 
The first part follows from this sequence of equalities. 
\[\textstyle 
\left(\frac{\prod_i\prod_j v_{ij}^{b_{ij}}}{\prod_j q_j^{q_j}}\right)^{1/n} = 
\left(\prod_i\prod_{j\in S_i} \frac{v_{ij}^{b_{ij}}}{b_{ij}^{b_{ij}}}\right)^{1/n} =
\left(\prod_i\prod_{j\in S_i} \left(\sum_{k\in S_i} v_{ik}\right)^{b_{ij}}\right)^{1/n} =
\left(\prod_i \sum_{k\in S_i} v_{ik}\right)^{1/n}
\]

For the rest of the proof, we work with the $\log$ transformation of the objective. Given the 
integral choices, the \ref{SR} program decomposes into a sum of separate  mathematical programs, one for each buyer $i$. 
\[ \textstyle\max \sum_{j\in S_i} \left( \bij \log \vij -  \bij\log \bij\right)  \text{ s.t.} \]
\[ \textstyle \forall i, \sum_{j\in S_i} \bij = 1,\text{and}~~ \forall i, j\in S_i, ~~~ \bij \geq 0 . \]
This is the same as minimizing the relative entropy, or KL-divergence, between two probability distributions, 
where the $\bij$s form one probability distribution, and the other distribution is given by $\frac \vij {\sum_{k \in S_i} v_{ik}}$. 
By Gibbs' inequality, it is known that this is minimized when the two distributions are the same, i.e., when
$b_{ij}=\frac \vij {\sum_{k \in S_i} v_{ik}}$. 
(We give an alternate proof of Gibbs' inequality using convex program duality in \Cref{app:convexduality}.)
\end{proof}
\subsection{Relaxation of the SR program}\label{sec.relaxation}
In designing their approximation algorithm for the NSW problem in~\cite{CG15},
they used, as an intermediate step, a fractional allocation, which 
was the equilibrium of a spending-restricted market with $\SB_j=1$ for all $j$.
If the price of an item $j$ is $p_j$, then this constraint could be expressed as $\sum_i x_{ij}p_j\leq 1$.
But, they could not introduce this constraint into the EG program, since it
combines both the primal variables $x_{ij}$ and the dual variable $p_j$. In the
absence of a program that could compute this fractional solution, they instead 
had to propose a complicated market equilibrium computation algorithm. Lemma~\ref{lem:fsreq} 
shows that in the \ref{SR} program, once we drop the constraint 
that $b_{ij}\in\{0, q_j\}$, the relaxed program, f-SR, computes the SR equilibrium. 
Unlike the EG program, the constraint that the total spending on any given item 
is at most $1$ involves only the primal variables $q_j$. If we also apply a logarithmic 
transformation to the objective function, then we get the convex program ($\CCP$)
of Section \ref{sec.4}, with the additional constraint that $q_j\leq 1$.
As a result, we provide a simple convex program that can compute the SR equilibrium. 
The proof of the following lemma essentially shows that the complementary 
slackness conditions are equivalent to market equilibrium conditions. 



\begin{restatable}{lemma}{fsreq}\label{lem:fsreq}
The f-SR program computes the SR equilibrium. The variables $\bij$ capture the amount of money spent by buyer $i$ on good $j$, 
and the variables $q_j$ capture the total spending on good $j$. The prices $p_j$ can be recovered from the optimal dual variables.  
\end{restatable}
\suppress{
\begin{proof}\nikhil{May be move this proof also to the Appendix.}
Let $\lambda_j, \mu_j, \eta_i$ be the dual variables corresponding to the first
three constraints of the SR program. By the KKT conditions, optimal solutions 
must satisfy the following: \vasilis{We may want to compress this a bit.}
\begin{enumerate}
\item $\forall i \in B,j \in A: \quad \log u_{ij} - \lambda_j  - \eta_i \leq 0, \quad b_{ij} >0 \Rightarrow \log u_{ij} - \lambda_j - \eta_i = 0, 
\quad - \log q_j + \lambda_j - \mu_j= 0$
\item $\forall j \in A : \quad  \mu_j \geq 0, \quad  \mu_j > 0 \Rightarrow q_j =  \SB_j $
\end{enumerate}

From 1, we have $\forall i \in B,j \in A$ : $ \frac{u_{ij}}{q_j e^{\mu_j}} \leq e^{\eta_i}$
and if $b_{ij} > 0$ then $\frac{u_{ij}}{q_j e^{\mu_j}} = e^{\eta_i}.$
Let $p_j = q_j e^{\mu_j}$. We will show that $p$ is an equilibrium price with spending $b$. From the above observation, it is easy to see that each buyer $i$ only spends money on his maximum bang-per-buck (MBB) goods at price $p$, i.e., goods that give her maximum utility per unit money spent. We also have to check that an optimal solution given by the convex program satisfies the market clearing conditions. The constraint that $\sum_j b_{ij}=1$ guarantees that each buyer $i$ must spend all his money. Therefore, we only have to show that the amount seller $j$ earns is the minimum between $p_j$ and $\SB_j$. If $q_j = \SB_j$ and $q_j \leq q_j e^{\mu_j} = p_j$. If $q_j < \SB_j$ then $\mu_j = 0$ and $p_j = q_j < \SB_j$. Thus, in both cases, $q_j = \min (p_j ,\SB_j)$ as desired.
\end{proof}
}

\paragraph{Existence and uniqueness of the SR equilibrium:} 

We study existence and uniqueness of the SR equilibrium in Appendix \ref{existence}.
We show an SR equilibrium exists if and only if $\sum_j \SB_j \geq \sum_i B_i$. On the uniqueness side, we 
show that the spending vector $q=(q_1,\ldots,q_m)$, where $q_j$ is the money spent on good $j$, is unique. 
Although in the Fisher model we have the uniqueness of price equilibrium, it is easy to see that this is not 
true for the SR equilibrium. Consider a market with only one buyer with utility function $u(x)=x_1$ and one seller. 
Let $B_1=1$ and $\SB_1=1$. It is easy to see every price bigger than 1 is an SR equilibrium price.  

\paragraph{Relation to \srub:}Quite surprisingly, we also show that the optimal objective value 
of the f-SR program is the same, up to scaling of the valuations, as the upper bound used by \cite{CG15}, which we called $\srub$. 

\begin{lemma}\label{lem:fsrub}
	The optimal value of the f-SR program is equal to \srub. 
\end{lemma}
\begin{proof}
	Let $\bar{b}_{ij}$ and $\bar{q}_j$ be an optimum solution to the f-SR program, and $\bar{x}$ and $\bar{p}$ be equilibrium allocation and price vectors resp.
	From \Cref{lem:fsreq}, the relation between these is that 
	$\bar{b}_{ij} = \bar{p}_j\bar{x}_{ij}$ and $\bar{q}_j = \min\{1,\bar{p}_j\}.$
	Recall that, from the definition of \srub,  we normalize each agent's valuations so that $v_{ij} =\bar{p}_j$ if $\bar{x}_{ij}>0$. 
	With this scaling of the valuations, the objective function of the f-SR program becomes
\[\textstyle \left(\frac{\prod_i\prod_j v_{ij}^{\bar{b}_{ij}}}{\prod_j \bar{q}_j^{\bar{q}_j}}\right)^{1/n} ~=~
\left(\frac{\prod_j \bar{p}_{j}^{\sum_i \bar{b}_{ij}}}{\prod_j \bar{q}_j^{\bar{q}_j}}\right)^{1/n} ~=~
\left(\prod_j \left(\bar{p}_j/\bar{q}_j\right)^{\bar{q}_j}\right)^{1/n} ~=~ \left(\prod_{j:\bar{p}_j\geq 1} \bar{p}_j\right)^{1/n},\]
where in the last equality, we used the fact that $\bar{p}_j=\bar{q}_j$ if $\bar{p}_j<1$ and $\bar{q}_j=1$ otherwise.
\end{proof}
\vspace{-4 mm}
\paragraph{The SR program integrality gap:}
%
Given Lemmas~\ref{lem:SR_OPT}, \ref{lem:fsreq}, and \ref{lem:fsrub}, a lower bound on the integrality gap of 
the \ref{SR} program also implies a lower bound on the best approximation factor that
one can show by rounding a solution to f-SR, and comparing the objective obtained to \srub.  The next
lemma provides such a lower bound for the integrality gap.\vspace{-2 mm}
\begin{lemma}
The integrality gap of the program above is at least $e^{1/e}\approx 1.44$.
\end{lemma}\vspace{-4 mm}
\begin{proof}
Consider an instance with $n$ bidders and $m=(1+f)n$ items, where $f\in (0,1)$ is a constant. Each agent $i$ has a value of 0
for the first $n$ items, except item $i$, for which his value is $(1-f)$. The value of every agent for items
$n+1$ to $m$, hence referred to as the ``valuable'' items, is equal to $V$, which is much higher than 1. In the 
SR equilibrium for this instance, the prices will be $(1-f)$ for the first $n$ items and $V$ for the rest. Each 
agent $i$ will be spending $(1-f)$ of his budget on item $i$ and the remaining budget of $f$ on the valuable items.

The objective value for this fractional solution would therefore be equal to $V^f$. On the other hand, any
integral allocation would have to assign each one of the valuable items to a distinct agent, so the optimal
NSW would be $(1-f)^{1-f}\cdot (1-f+V)^f$. If we let $V$ go to infinity, this leads to an integrality gap of
\[\textstyle \lim_{V\to \infty}\left(\frac{V^f}{(1-f)^{1-f}\cdot (1-f+V)^f}\right) ~=~ \frac{1}{(1-f)^{1-f}}\]
which, for $f=(e-1)/e$, yields the desired $e^{1/e}$ integrality gap\footnote{To be precise, to make sure that
$m$ is an integer, $fn$ would also have to be an integer. Therefore, we as we let $n$ be arbitrarily large,
$f$ can take values arbitrarily close to $(e-1)/e$ while $fn$ remains an integer.}.
%
\end{proof}

\section{A Tight Analysis of the Spending-Restricted Rounding Algorithm}\vspace{-2 mm}
Using the SR equilibrium as a starting point,~\cite{CG15} proposed the a
rounding algorithm called the \emph{Spending-Restricted Rounding} (SRR) algorithm.
Using {\srub} as an upper bound, they showed that the approximation factor of this
algorithm is at most $2e^{1/e}\approx 2.89$. 
The first step of the SRR algorithm is to compute the SR equilibrium which,
in light of the previous section's results, we can now do using the f-SR
convex program. Then, for each tree of the spending graph $Q(b)$, it chooses
an arbitrary agent as the root and assigns all items that are either leaves
or have $q_j\leq 1/2$ to their parent-agent. The remaining items are matched 
to agents using the matching with the optimal NSW value, given the previous
assignments. This matching can be computed in polynomial using a maximum weight
matching algorithm and $\log v_{ij}$ as weights instead of $v_{ij}$ (see~\cite{CG15}
for more details). The (full) proofs of this section are deferred to Appendix~\ref{app:approx}.

\LinesNumbered	
\begin{algorithm}
\SetEndCharOfAlgoLine{.}
Compute a spending-restricted equilibrium $(b, q)$\;
Choose a root-agent for each tree in the spending graph $Q(b)$\;
Assign any leaf-item in the trees to its parent-agent\;
Assign any item $j$ with $q_j \leq 1/2$ to its parent-agent\;
Compute the optimal matching of the remaining items to adjacent agents\;
	\caption{Spending-Restricted Rounding (SRR) \cite{CG15}.}\label{alg3}
\end{algorithm}

Using a careful analysis, we now show that the approximation factor
of the SRR algorithm is, in fact, better than $2.89$ by proving an upper
bound of 2. We conclude this section with a matching lower bound.\vspace{-2 mm}

\begin{theorem}\label{thm:approx}
The approximation factor of the SRR algorithm is at most 2.
\end{theorem}
\vspace{-2 mm}
\begin{proof}[Proof Sketch]
For each item $j$ that has more than one child-agent in the spending 
graph $Q(b)$, remove the edges connecting it to all but the one child-agent 
that spends the most money on $j$, i.e., the one with the largest $b_{ij}$ value. 
This yields a pruned spending graph $P(b)$ that is also a forest of trees.
We refer to the trees of the pruned graph $P(b)$ as the {\em matching-trees}. 
In every matching-tree $T$ with $k\geq 2$ agents, when the algorithm
reaches its last step, every remaining item has exactly one parent-agent 
and one child-agent, so all but one agent can be matched to one of these
items. Our proof shows that there exists a matching of the remaining items 
such that the agents within $T$ have a ``high'' NSW.

A naive way to match the agents in the last step of the algorithm would be 
to match all of them, except the one that has accrued the highest value
during the previous steps. It was already observed in~\cite{CG15} that, 
for any matching-tree $T$ of $k$ agents, there exists an agent who was 
assigned value at least $1/(2k)$ during Steps 3 and 4 of the algorithm,
so we could match every agent in $T$, except him. But, what is the worst
case distribution of value that can arise in this matching? We show that
the worst case arises for matching-trees that contain a single agent and
no items with $p_j>1/2$. But, even in this case, such an agent got all 
the items that he was spending on in the SR equilibrium, except one, and
he could not be spending more than half of his budget on the one he lost.
To verify this fact, note that he either lost this item because the total 
money spent on the item was less than half, i.e., $q_j\leq 1/2$,  
and it was assigned to its parent at Step 4, or
because the edge connecting him to this item was pruned in the transition
from $Q(b)$ to $P(b)$. But, in both of these cases, he could not be
spending more than $1/2$ on that item, so he got at least half of his 
SR equilibrium value.

The more demanding part of the proof is to show that the worst case arises for
matching-trees of size $1$. In contrast to the analysis of~\cite{CG15}, we use
the vital observation that, if the agent of some matching-tree $T$ who does 
not get matched to an item has value $v_{\alpha}$, then every other agent $i\in T$
gets value at most $v_{ij}+v_{\alpha}$, 
where $j$ is the item that he was matched to in the last step.
Lemma~\ref{lem:heavy_agents} uses this fact to prove that in the worst
case distribution of value, at least 
$\left\lfloor\frac{k-2v_{\alpha}}{1+2v_{\alpha}}\right\rfloor$ 
agents get value greater than, or equal to, $1$. In other words, this
new lemma shows that, if the unmatched agent were to leave a lot of value 
on the table, then this value would not end up with just a few agents but, 
rather, it would have to be well distributed among the remaining agents.
Building further on this observation, Lemma~\ref{ineq:bound1} shows
that, for any matching-tree $T$ with $k$ agents, the allocation $x'$ 
induced by the naive matching algorithm satisfies
\[\textstyle \prod_{i\in T} v_i(x') ~\geq~ \frac{1}{2^k} \prod_{j\in T:p_j\geq 1} p_j.\]

Since the allocation $x$ that the SRR algorithm outputs is at least
as good as the one by the naive matching, we can combine this inequality 
with the $\srub$ upper bound to get the desired approximation factor bound:
\[\textstyle \left(\prod_i v_i(x)\right)^{1/n} ~=~ \left( \prod_T \prod_{i\in T} v_i(x)\right)^{1/n} ~\geq~ \frac{1}{2} \left(\prod_{j: p_j \geq 1} p_j \right)^{1/n}.\]\end{proof} \vspace{-2 mm}
\begin{lemma}\label{thm:approx_lb}
The approximation factor of the SRR algorithm is exactly $2$.
\end{lemma}\vspace{-4 mm}

\section{Discussion}\label{sec:Discussion}\vspace{-2 mm}
Regarding additional Fisher market extensions, an obvious open question is to obtain a convex program for the 
common generalization of the spending-restricted and utility-restricted markets, in which buyers have utility 
bounds and sellers have earning bounds, for the case of linear utilities.

Regarding the NSW problem, we have addressed the symmetric case of NSW, which assumes that all agents have equal 
budget (or clout). While introducing the Nash bargaining problem \cite{nash.bargain}, Nash only considered the 
symmetric case but, soon after that, Kalai proposed the non-symmetric case as well, which is also well-studied. 
Hence a natrual open problem is to obtain a constant factor approximation algorithm for the non-symmetric
case of NSW. The objective in this generalization is to maximize
$ \textstyle  \left(\prod_i u_i^{B_i} \right)^{1/B},$
where $B_i$ is the budget of agent $i$ and $B = \sum_i B_i$.
Another important generalization of NSW would be to consider  utilities that are subadditive instead of
additive. In particular, the case of submodular utilities would definitely deserve more attention.

\bibliographystyle{plainnat}
\bibliography{kelly} 

\newpage
\appendix
\section{Convex Programming Duality}
\label{app:convexduality}

\paragraph{Properties of $f^*$:} We note some useful properties of $f^*$ here. 
\begin{itemize}
  \item[$\bullet$] $f^* $ is strictly convex and differentiable. (even if $f$ is not strictly convex and differentiable)
  \item[$\bullet$] $f^{**} = f$. (using the assumption that $f$ is strictly convex and differentiable)
  \item[$\bullet$] If $f$ is separable, that is $f(x) = \sum_i f_i(x_i)$, then $f^*(\mu) =\sum_i f_i^*(\mu_i)$.
  \item[$\bullet$] If $g(x) = cf(x)$ for some constant $c$, then $g^*(\mu) = c f^*(\mu/c)$.
  \item[$\bullet$] If $g(x) = f(cx)$ for some constant $c$, then $g^*(\mu) = f^*(\mu/c)$.
  \item[$\bullet$] If $g(x) = f(x+ a)$ for some constant $a$, then $g^*(\mu) = f^*(\mu) - \mu^T a$.
  \item[$\bullet$] If $\mu$ and $x$ are such that $f(x) + f^*(\mu) = \mu^Tx$ then
$\grad f(x) = \mu $ and $\grad f^*(\mu) = x$.
\item[$\bullet$] Vice versa, if $\grad f(x) = \mu $ then $\grad f^*(\mu) = x$ and
$f(x) + f^*(\mu) = \mu^Tx$.
\end{itemize}

\paragraph{Conjugates of some simple strictly convex and differentiable functions} 
\begin{itemize}
	\item[$\bullet$] If $f(x) = \frac{1}{2} x^2$, then $\nabla f (x) = x$. Letting $\mu = x$ in $\mu^Tx - f(x)$, leads to $f^*(\mu) =\frac{1}{2} \mu^2$.
	\item[$\bullet$] If $f(x) = -\log(x)$, then $\nabla f (x) = \frac{-1}{x}$. Set $\mu = \frac{-1}{x}$ to get $f^*(\mu) = -1 + \log(x) = -1 - \log(-\mu)$.
	\item[$\bullet$] If $f(x) = x \log x$, then $\nabla f(x) = \log x + 1 = \mu$. So $x = e^{\mu -1}$.
	$f^*(\mu) = \mu x - f(x) = x (\log x +1 ) - x \log x = x = e^{\mu-1}$.
	That is, $f^*(\mu)  = e^{\mu-1}$.
	
	\end{itemize}
	
	\CPduality*
	\begin{proof}
		
		Suppose first that the set of linear constraints is itself infeasible, that is, there is no solution to the set of inequalities 
		\begin{equation}\label{eq.feasible} \forall~ j, \sum_i \aij x_i \leq b_j.\end{equation}
		Then by Farkas' lemma, we know that there exists numbers $\lambda_j\geq 0$ for all $j$ such that 
		\[ \forall~ i, \sum_j \aij \lambda _j = 0, \text{and} \sum_j \lambda_j b_j < 0.\]
		Now consider the dual solution with these $\lambda_j$s and $\mu_i = c_i$. 
		This is feasible, and the dual objective is $ f^*(c) +  \sum_j \lambda_j b_j $. 
		By multiplying all the $\lambda_j$s by a large positive number, the dual objective can be made arbitrarily small (goes to $-\infty$). 
		
		Now suppose that the feasible region defined by the inequalities (\ref{eq.feasible}) and the domain of $f$ defined 
		as $ dom(f) = \{x: f(x) < \infty \}$ are disjoint. Further assume for now that $f^*(c) < \infty$ and that 
		there is a strict separation between the two, meaning that for all $x $ feasible and $y \in dom(f)$,  
		$d(x,y) > \epsilon$ for some $\epsilon > 0$. 
		Then once again by Farkas' lemma we have that there exist $\lambda_j\geq 0$ for all $j$ and $\delta > 0$ 
		such that 
		\[\forall y \in dom(f),   \sum_{i,j} \aij \lambda _j y_i  >  \sum_j \lambda_j b_j (1 + \delta) .\]
		This implies that the dual objective is $< f^*(c) -  \delta \sum_j \lambda_j b_j $, and as before, 
		by multiplying all the $\lambda_j$ by a large positive number,  $g$ can be made arbitrarily small.

		Now we may assume that the primal is feasible. Define the Lagrangian function
		\[L(x,\lambda) :=  \sum_i c_i x_i - f(x) +\sum_j \lambda_j
		(b_j -  \sum_i \aij x_i ).\]
		We say that $x$ is feasible if it satisfies
		all the constraints of the primal problem.
		Note that for all $\lambda \geq 0 $ and $x$ feasible,
		$L(x,\lambda) \geq \sum_i c_i x_i - f(x) $.
		Define the dual function
		\[ g(\lambda) = \max_x L(x,\lambda).\]
		So for all $\lambda,x$,  $g(\lambda) \geq L(x,\lambda).$
		Thus
		$\min_{\lambda\geq 0} g(\lambda) $ is an upper bound on the optimum value for the primal program.
		The dual program is essentially $\min_{\lambda\geq 0} g(\lambda)$. We further simplify it as follows.
		Letting  $\mu_i = c_i - \sum_j \aij \lambda_j,$ we can rewrite the expression for $L$ as
		\[L = \sum_i \mu_i x_i - f(x) + \sum_j b_j \lambda_j.\]
		Now note that $g(\lambda) = \max_x L(x,\lambda) = \max_x \{\sum_i \mu_i x_i - f(x)\} + \sum_j b_j \lambda_j = f^*(\mu) + \sum_j b_j \lambda_j$.
		Thus we get the dual optimization problem:
		\[\min \textstyle  \sum_j b_j\lambda_j + f^*(\mu) \text{ s.t.} \]
		\[\forall~ i, \textstyle \sum_j \aij \lambda_j  = c_i - \mu_i,\]
		\[\forall~ j, \lambda_j \geq 0 .\]
	\end{proof}

	\EGdual* 
	\begin{proof}
		We let the dual variable corresponding to the constraint $ u_i \leq \sum_j \uij \xij$ be $\beta_i$ and 
		the dual variable corresponding to the constraint $ \sum_i \xij \leq 1$ be $p_j$.  
		We also need a variable $\mu_i$ that corresponds to the variable $u_i$ in the primal program since it appears in the 
		objective in the form of a  concave function, $m_i \log u_i$. We now calculate the conjugate of this function. 
		Recall that if $f(x) = -\log x$ then $f^*(\mu) = -1 - \log (-\mu)$,
		and if $g(x) = c f(x)$ then $g^*(\mu) = c f^*(\mu/c)$.
		Therefore if $g(x) = - c \log x$ then $g^*(\mu) = -c - c\log (-\mu/c) =
		c \log c - c -c\log (-\mu)$. In the dual objective, we can ignore the constant terms,
		$c \log c - c$. We are now ready to write down the dual program which is as follows.
		\[\textstyle \min  \sum_j p_j  - \sum_i m_i \log (-\mu_i) \text{ s.t.}\]
		\[\textstyle \forall~i,j, p_j \geq \uij\beta_i,\]
		\[ \forall~i, \beta_i = - \mu_i  .\]
		We can easily eliminate $\mu_i$ from the above to get the program as stated in the lemma.
	\end{proof}

	\Shmyrev* 
	\begin{proof}
We construct the dual of (\ref{cp.dualeglogs}) as outlined in the \Cref{sec:CPduality}. 
Again, we need to calculate the conjugate of the convex function that appears in the objective, namely $e^x$. We could calculate it from scratch, or derive it from the ones we have already calculated. 
Recall that if $f(x) = e^{x-1}$, then $f^*(\mu) = \mu \log \mu$,
and if $g(x) = f(x+a)$ then $g^*(\mu) = f^*(\mu) - \mu^T a$. Thus if $g(x) = e^x = f(x+1)$ then
$g^*(\mu)= f^*(\mu) - \mu = \mu \log \mu - \mu. $ The dual variable corresponding to the constraint
$\gamma_i + q_j  \geq \log \uij$ is $\bij$ and the dual variable corresponding to $e^{q_j}$ is $p_j$.
The structure of the dual program now follows from \Cref{lem:CPduality}. 
\end{proof}

\subsection{Extensions}

The Eisenberg-Gale convex program can be generalized to capture the equilibrium of many other markets, such as markets
with Leontief utilities, or network flow markets. In fact,  \cite{JVEG} identify a whole class of such markets
whose equilibrium is captured by convex programs similar to that of Eisenberg and Gale (called {\em EG markets}). We can take the dual
of all such programs to get corresponding generalizations for the convex program (\ref{cp.dualeg}).
For instance, a Leontief utility is of the form $U_i = \min_j \left\{ \xij/\phi_{ij} \right \}$ for some given values $\phi_{ij}$. 
The Eisenberg-Gale-type convex program for Fisher markets with Leontief utilities is as follows, 
along with its dual (after some simplification as before). 

\begin{minipage}{0.35\textwidth}  
	\[ \textstyle \text{Primal: }\max \sum_{i} m_i \log u_i \text{ s.t.} \]
	\[ \textstyle\forall~ i,j,  u_i \leq  \xij/\phi_{ij} ,\]
	\[ \textstyle\forall~j, \sum_i \xij \leq 1, \]
	\[ \xij \geq 0.\]
\end{minipage}
\hfill\vline\hfill
\begin{minipage}{0.35\textwidth}  
	\[\textstyle \text{Dual: }\min \sum_j p_j  - \sum_i m_i \log (\beta_i) \text{ s.t.}\]
	\[ \textstyle\forall~i, \sum_j \phi_{ij} p_j = \beta_i.\]
\end{minipage}

In general for an EG-type convex program, the dual has the objective function $ \sum_j p_j  - \sum_i m_i \log (\beta_i)$
where $\beta_i$ is the minimum cost buyer $i$ has to pay in order to get one unit of utility. 
For instance, for the network flow market, where the goods are edge capacities in a network and the buyers are source-sink 
pairs looking to maximize the flow routed through the network, then $\beta_i$ is the cost of the cheapest path between the 
source and the sink given the prices on the edges. 

However, for some markets, it is not clear how to generalize the Eisenberg-Gale convex program, 
but the dual generalizes easily.  In each of the cases, the optimality conditions can be easily seen to be 
equivalent to equilibrium conditions. We now show some examples of this.
\subsubsection*{Quasi-linear utilities}
Suppose the utility of buyer $i$  is $\sum_j (\uij - p_j)\xij$.
In particular, if all the prices are such that $p_j > \uij$, then the buyer prefers to not be allocated any good and go
back with his budget unspent. It is easy to see that the following convex program captures the equilibrium prices for such utilities. 
In fact, given this convex program, one could take its dual to get an EG-type convex program as well. 

\begin{minipage}{0.35\textwidth} 
	\begin{equation}\label{cp.dualegQuasilinear1}
		\textstyle \text{Primal: }\min \sum_j p_j  - \sum_i m_i \log (\beta_i) \text{ s.t.}
	\end{equation}
	\[\textstyle \forall~i,j, p_j \geq \uij\beta_i,\]
	\[ \forall~i, \beta_i \leq 1. \]
\end{minipage}
\hfill\vline\hfill
\begin{minipage}{0.35\textwidth} 
	\[ \textstyle \text{Dual: }\max \sum_{i} m_i \log u_i -v_i \text{ s.t.} \]
	\[\textstyle \forall~ i, u_i \leq \sum_j \uij \xij + v_i ,\]
	\[ \textstyle\forall~j, \sum_i \xij \leq 1, \]
	\[ \xij ,v_i \geq 0 .\]
\end{minipage}

Although this is a small modification of the Eisenberg-Gale convex program, it is not clear how one would arrive at this 
directly without going through the dual. 
\subsubsection*{Transaction costs}
Suppose that we are given, for every pair, buyer $i$ and good $j$, a transaction cost $\cij$ that the buyer has to pay per unit
of the good in addition to the price of the good. Thus the total money spent by buyer $i$ is $\sum_j (p_j + \cij) \xij$.
\citet{CDK10} show that the following convex program captures the equilibrium prices for such markets.
\begin{equation}\label{cp.dualegQuasilinear2}\textstyle \min \sum_j p_j  - \sum_i m_i \log (\beta_i) \text{ s.t.}\end{equation}
\[ \textstyle\forall~i,j, p_j + \cij \geq \vij\beta_i,\]
\[\textstyle \forall~i, \beta_i \leq 1. \]

\subsubsection*{Alternate proof of Gibbs' inequality}
Consider the following convex program. 
\[ \textstyle\max \sum_{j\in S_i} \left(\bij \log \vij -  \bij\log \bij\right)  \text{ s.t.} \]
\[ \textstyle \sum_{j\in S_i} \bij = 1,\]
\[\bij \geq 0 ~~\forall j\in S_i. \]
Using the duality techniques developed in this paper, we write the following dual of this program. 
\[ \textstyle \min \alpha_i + \sum_{j \in S_i}  e^{\muij -1} \text{s.t.} \]
\[ \textstyle  \forall j \in S_i, \alpha_i \geq \log \vij - \muij. \]
Suppose that we fix the value of  $\alpha_i$. 
Given this, we want to set $\muij$ to be as small as possible s.t.\ the constraint 
$\alpha_i \geq \log \vij - \muij$ is satisfied, which gives us $\muij = \log \vij - \alpha_i$. 
Then $e^{\muij -1} = \vij e^{-1-\alpha_i}$, and the objective can be written as a function of $\alpha_i$ as 
\[  \alpha_i + \sum_{j \in S_i } \vij e^{-1-\alpha_i}.\]
This can be minimized by setting the derivative to zero, which gives 
\[ 1 - \sum_{j \in S_i} \vij e^{-1-\alpha_i}= 0\]
\[ \Leftrightarrow e^{\alpha_i +1} = \sum_{j \in S_i} \vij \Leftrightarrow \alpha_i +1 = \log (\sum_{j\in S_i} \vij) .\]
The minimum value of the objective is then $\alpha_i +1 = \log (\sum_{j\in S_i} \vij) $, 
which is also obtained in the primal by setting $b_{ij}=\frac{v_{ij}}{\sum_{k \in S_i} v_{ik}}$. 
 
\section{Convex Program, Existence and Uniqueness for the SR equilibrium}\label{existence}

In this section, we give the proof of \Cref{lem:fsreq}, that the f-SR program captures the SR equilibrium. 
We then study the existence and the uniqueness of the SR equilibrium and we
show a necessary and sufficient condition for its existence. On the uniqueness side, we 
show that the spending vector $q=(q_1,\ldots,q_m)$, where $q_j$ is the money spent on good $j$, is unique. 
Although in the Fisher model we have the uniqueness of price equilibrium, it is easy to see that this is not 
true for the SR equilibrium. Consider a market with only one buyer with utility function $u(x)=x_1$ and one seller. 
Let $B_1=1$ and $\SB_1=1$. It is easy to see that every price bigger than 1 is an SR equilibrium price.  

We first state the f-SR program, with a log transformation of the objective function, and generalized for arbitrary spending limits for each good, as in the definition of the general SR equilibrium model. 
This convex program is a natural extension of program $\mathcal{CP}$ presented in Section \ref{sec.4}, 
with an additional set of constraints for sellers having earning limits:
\[ \textstyle \max \sum_{i,j} b_{ij} \log v_{ij} - \sum_{j} ( q_j \log q_j - q_j ) ~ \text{s.t.} \tag{f-SR} \label{cp4} \]
\begin{equation}\label{con1.1}
\textstyle
\forall j,  \sum_i b_{ij} = q_j  ,
\end{equation}
\begin{equation}\label{con1.3}
\textstyle
\forall i,\sum_j b_{ij} = B_i  ,
\end{equation}
\begin{equation}\label{con1.2}
\textstyle
\forall j,q_j \leq \SB_j ,
\end{equation}
\begin{equation}\label{con1.4}
\textstyle
\forall i,j,b_{ij} \geq 0.
\end{equation}
Here $b_{ij}$ is the amount of money buyer $i$ spends on good $j$, and $q_j$ is the total amount of spending on good $j$. Constraint \ref{con1.2} makes sure that the spending on good $j$ does not exceed the earning limit of seller $j$. 

\fsreq*
	\begin{proof}
		Let $\lambda_j, \mu_j, \eta_i$ be the dual variables corresponding to the first
		three constraints of the SR program. By the KKT conditions, optimal solutions 
		must satisfy the following: 
		\begin{enumerate}
			\item $\forall i \in B,j \in A: \quad \log v_{ij} - \lambda_j  - \eta_i \leq 0$
			\item $\forall i \in B,j \in A: \quad b_{ij} >0 \Rightarrow \log v_{ij} - \lambda_j - \eta_i = 0$
			\item $\forall j \in A : \quad - \log q_j + \lambda_j - \mu_j= 0$
			\item $\forall j \in A : \quad  \mu_j \geq 0$
			\item $\forall j \in A : \quad  \mu_j > 0 \Rightarrow q_j =  \SB_j $
		\end{enumerate}
		
		From the first 3 conditions, we have $\forall i \in B,j \in A$ : $ \frac{v_{ij}}{q_j e^{\mu_j}} \leq e^{\eta_i}$
		and if $b_{ij} > 0$ then $\frac{v_{ij}}{q_j e^{\mu_j}} = e^{\eta_i}.$
		Let $p_j = q_j e^{\mu_j}$. We will show that $p$ is an equilibrium price with spending $b$. From the above observation, it is easy to see that each buyer $i$ only spends money on his maximum bang-per-buck (MBB) goods at price $p$, i.e., goods that give her maximum utility per unit money spent. We also have to check that an optimal solution given by the convex program satisfies the market clearing conditions. The constraint that $\sum_j b_{ij}=1$ guarantees that each buyer $i$ must spend all his money. Therefore, we only have to show that the amount seller $j$ earns is the minimum between $p_j$ and $\SB_j$. If $q_j = \SB_j$ and $q_j \leq q_j e^{\mu_j} = p_j$. If $q_j < \SB_j$ then $\mu_j = 0$ and $p_j = q_j < \SB_j$. Thus, in both cases, $q_j = \min (p_j ,\SB_j)$ as desired.
	\end{proof}

\begin{lemma} \label{earning_existence}
An SR equilibrium price exists if and only if $\sum_j \SB_j \geq \sum_i B_i$. 
\end{lemma}
\begin{proof} An equilibrium price exists if and only if the feasible region of the f-SR convex program is not empty. 
We first prove that for the case of linear utility function, the program is feasible if and only if 
$\sum_j \SB_j \geq \sum_i B_i$. If $\sum_j \SB_j < \sum_i B_i$ then the feasible region is empty because the 
set of constraints \ref{con1.1}, \ref{con1.2} and \ref{con1.3}  can not be satisfied together. If 
$\sum_j \SB_j \geq \sum_i B_i$ then $y_{ij} = \frac{B_i \SB_j}{ \sum_j \SB_j}$ gives a feasible solution 
because $\sum_i y_{ij} = \SB_j \frac{\sum_i B_{i}}{\sum_j \SB_j} \leq \SB_j$ and $\sum_j y_{ij} = B_i \frac{\sum_j \SB_{j}}{\sum_j \SB_j} = B_i$. 
\end{proof}

\begin{lemma}\label{uniqueness}
The spending vector $q$ of the SR equilibrium is unique.
\end{lemma}
\begin{proof}
Consider two distinct price equilibria $p$ and $p'$, their corresponding spending vectors $q$ and $q'$ and their corresponding demand vectors $x$ and $x'$.
Note that $p_j \geq p'_j \Rightarrow q_j\geq q'_j$ because $q_j=x_j p_j= \min (1,\frac{\SB_j}{p_j})p_j\geq \min (1,\frac{\SB_j}{p'_j})p'_j=q'_j$. 
Consider price vector $r=(r_1,\ldots,r_m)$ where $\forall k,$ $r_k= \max (p_k,p'_k)$, its corresponding spending vector $q^r$ and its corresponding demand vectors $x^r$.
Note that by changing prices from $p$ to $r$ we may only increasing the prices. Therefore, it is easy to see under linear utility functions the demand of good $j$ going from prices $p$ to $r$ would not decrease if $p'_j<p_j=r_j$. 
%
Therefore, 
we have $ q^r_j=x^r_j r_j=x^r_j p_j\geq x_j p_j=q_j\geq q'_j$. We can do the same for all $j$ and show $\forall j,$ $q^r_j=max(q_j,q'_j)$.
For the sake of a contradiction suppose $\exists j$, $q_j>q'_j$ then using the later it is easy to show $\sum_j q^r_j > \sum_j q_j = \sum_j q'_j=\sum_i B_i$ which is contradiction because the money spent on goods cannot be more than the total budget. Therefore, $\forall j$ , $q_j=q'_j$ and the lemma follows.
\end{proof}
\subsection{Rationality of the SR equilibrium}
\label{sec.rationality}

In this section, we prove rationality results for the spending restricted outcome. 
Specifically, we show that for those market models, a rational equilibrium exists 
if an equilibrium exists and all the parameters are rational numbers.

\begin{lemma} \label{earning_rational}
In spending-restricted market model under linear utility functions, a rational equilibrium exists if $\sum_j \SB_j \geq \sum_i B_i$ and all the parameters specified are rational numbers. 
\end{lemma}
\begin{proof}
Let $A_i$ be the set of goods that buyer $i$ spends money on, $\mathcal{A}$ be the family of $A_i$'s, and $L$ be the set of sellers reaching their earning limits. An equilibrium price $p$, the corresponding spending $b$ and inverse MBB value $\alpha$, if existed, must be a point inside the polyhedron $P(\mathcal{A},L)$ bounded by the following constraints:
\begin{align*} 
\textstyle 
\forall i\in N, \forall j \in A_i \qquad &  v_{ij} \alpha_i =  p_j \\
\textstyle 
\forall j \in M \qquad & v_{ij} \alpha_i \leq p_j \\
\textstyle 
\forall i \in N, \forall j \not \in A_i \qquad & b_{ij} = 0  \\
\textstyle 
\forall i \in N, \qquad & \sum_j b_{ij} = B_i \\
\textstyle 
\forall j \in L \qquad & \sum_{i} b_{ij}= \SB_j \qquad p_j \geq \SB_j \\ 
\textstyle 
\forall j \not \in L \qquad & \sum_{i} b_{ij} = p_j \qquad p_j \leq \SB_j \\ 
\textstyle 
\forall i \in N, j \in M \qquad & b_{ij} \geq 0
\end{align*}
If an equilibrium price exists, then $\mathcal{A}$ and $L$ such that $P(\mathcal{A},L)$ is non-empty must also exist. Every point inside that non-empty polyhedron must also correspond to an equilibrium price. Since $v_{ij}$'s, $B_i$'s and $\SB_j$'s are rational numbers, a vertex of $P(\mathcal{A},L)$ gives a rational equilibrium price. It then follows from Lemma \ref{earning_existence} that a rational equilibrium exists if and only if $\sum_j \SB_j \geq \sum_i B_i$.
\end{proof}

\section{SR equilibrium with Spending Constraint Utilities}
\label{sec:m2sc} 
We next define the spending constraint model.
As before, let $M$ be a set of divisible goods and $N$ a set of buyers, $|M|=m, \ |N| = n$.
Assume that the goods are numbered from 1 to $m$ and the buyers are numbered
from 1 to $n$. 
Each buyer $i \in N$ comes to the market with a specified amount of
money, say $B_i \in \Qplus$, and we are specified the quantity, 
$b_j \in \Qplus$ of
each good $j \in M$. 
For $i \in N$ and $j \in M$, let $f_j^i: \ [0, B_i] \rightarrow \Rplus$
be the {\em rate function} of buyer $i$ for good $j$; it specifies the rate
at which $i$ derives utility per unit of $j$ received, as a function of
the amount of her budget spent on $j$. If the price of $j$ is fixed at
$p_j$ per unit amount of $j$, then the function $f_j^i/p_j$
gives the rate at which $i$ derives utility per dollar spent, as a function of
the amount of her budget spent on $j$. Define $g_j^i: [0, B_i] \ra \Rplus$ as follows:
\[ g_j^i(x) = \int_0^x { {\frac{f_j^i(y)}{p_j}} dy }  .\]
This function gives the utility derived by $i$ on spending $x$ dollars on good $j$ at price $p_j$.

In this paper, we will deal with the case that $f_j^i$'s
are decreasing step functions. If so, $g_j^i$ will be a piecewise-linear and concave
function. The linear version of Fisher's problem \cite{scarf} is the special
case in which each $f_j^i$ is the constant function so that $g_j^i$ is a linear function.
Given prices $\pp = (p_1, \ldots, p_m)$ of all goods, each buyer wants a utility maximizing bundle of goods.
Prices $\pp$ are equilibrium prices if each good with a positive price is fully sold.

The convex program for spending restricted model under spending constraint utility functions is as follows:
\begin{equation}\label{cp5}
\max \sum_{i,j,l} b^l_{ij} \log v^l_{ij} - \sum_{j} ( q_j \log q_j - q_j ) ~ \text{s.t.} \tag{P2}
\end{equation}
\begin{equation} \label{con5.1}
\forall j,\sum_{i,l} b_{ij}^l = q_j,
\end{equation}
\begin{equation}\label{con5.3}
\forall i, \sum_{j,l} b_{ij}^l = B_i,
\end{equation}
\begin{equation} \label{con5.4}
\forall i,j, l \in S,b_{ij}^l \leq  B_{ij}^l,
\end{equation}
\begin{equation}\label{con5.2} 
\forall j,q_j \leq \SB_j,
\end{equation}
\begin{equation} \label{con5.5}
\forall i,j, l\in S,b_{ij}^l \geq 0. 
\end{equation}
Here $b_{ij}^l$ is the amount of money buyer $i$ spends on good $j$ under segment $l$, $B_{ij}^l$ is length of the segment $l$, and $q_j$ is the total amount of spending on good $j$.

\begin{lemma} 
	\label{lem:cpm1sc}
	Convex program \ref{cp5} captures SR equilibrium prices of SR market model under spending constraint utility function.
\end{lemma}
\begin{proof}
	Let $\lambda_j, \mu_j, \eta_i, \gamma_{ijl}$ be the dual variables for constraints \ref{con5.1}, \ref{con5.2}, \ref{con5.3}, \ref{con5.4} respectively. By the KKT conditions, optimal solutions must satisfy the following:
	\begin{enumerate}
		\item $\forall i \in N,j \in M, l \in S: \quad \log v^l_{ij} - \lambda_j  - \eta_i - \gamma_{ijl} \leq 0$
		\item $\forall i \in N,j \in M, l \in S: \quad b^l_{ij} >0 \Rightarrow  \log v^l_{ij} - \lambda_j  - \eta_i - \gamma_{ijl} = 0$
		\item $\forall j \in M : \quad - \log q_j + \lambda_j - \mu_j= 0$
		\item $\forall j \in M : \quad  \mu_j \geq 0$
		\item $\forall j \in M : \quad  \mu_j > 0 \Rightarrow q_j =  \SB_j $
		\item $\forall i \in N,j \in M, l \in S: \quad \gamma_{ijl} \geq 0$
		\item $\forall i \in N,j \in M, l \in S: \quad \gamma_{ijl} >0 \Rightarrow b_{ij}^l =  B_{ij}^l  $
		\end{enumerate}
		
		Let $p_j = q_j e^{\mu_j}$. We will prove that $p$ is an equilibrium price with spending $b$. The second KKT condition says that for a fixed pair of buyer $i$ and good $j$, $b_{ij}^l > 0$ implies 
		\[ \frac{v_{ij}^l}{e^{\gamma_{ijl}}} = e^{\lambda_j} e^{\eta_i} \]
		Therefore, the ratio $v_{ij}^l/ e^{\gamma_{ijl}}$ is the same for every segment $l$ under which $i$ spends money on $j$. From KKT condition 7, $\gamma_{ijl} > 0$ implies $b_{ij}^l =  B_{ij}^l$. It follows that for each good $j$, $i$ must finish spending money on a segment with higher rate before starting spending money on a segment with lower rate. 
		
		From the first 3 KKT conditions, we have:
		\[ \frac{v_{ij}^l}{q_j e^{\gamma_{ijl}} e^{\mu_j}} \leq e^{\eta_i}\]
		and equality happens when $b_{ij}^l > 0$. For every segment that $i$ can still spend money on, $b_{ij}^l$ must be less than $B_{ij}^l$, and thus $\gamma_{ijl} = 0$. Therefore, for every $j$ and $l$ such that  $B_{ij}^l > b_{ij}^l > 0$, we have 
		\[ \frac{v_{ij}^l}{p_j} = \frac{v_{ij}^l}{q_j e^{\mu_j}} = e^{\eta_i}\]
		and this ratio $\frac{v_{ij}^l}{p_j}$ is maximized among all segments that $i$ can spend money on, i.e. segments such that $b_{ij}^l < B_{ij}^l$. Therefore, we can conclude that each buyer $i$ is spending according to his best spending strategy. 
		
		By complementary slackness condition, if $q_j  < \SB_j$ then $\mu_i = 0$ and $q_j = p_j$.  Otherwise, if $p_j = \SB_j$ then $q_j \leq p_j$. Therefore, in this model, the amount seller $j$ earns is the minimum between $\SB_j$ and $p_j$.
\end{proof}

\paragraph{Existence and Uniqueness}

We first show that the same condition that works for linear utilities also works for spending constraint utilities.

\begin{lemma} 
For spending constraint utility functions, an equilibrium price exists if and only if $\sum_j \SB_j \geq \sum_i B_i$. 
\end{lemma}
\begin{proof} An equilibrium price exists if and only if the feasible region of the convex program is not empty. 
Similarly to the proof of Lemma \ref{earning_existence}, we can prove that the program is feasible if and only if 
$\sum_j \SB_j \geq \sum_i B_i$. If $\sum_j \SB_j < \sum_i B_i$ then the feasible region is empty because the set 
of constraints \ref{con5.1}, \ref{con5.2} and \ref{con5.3} can not be satisfied together. Using a similar argument 
as in the previous part, we can show that if the amount of money that $i$ spends on $j$ is $B_i \SB_j /  \sum_j \SB_j$ 
then constraints \ref{con5.1}, \ref{con5.2} and \ref{con5.3} are all satisfied. We only need to guarantee that 
contraint \ref{con5.4} is satisfied as well. This can be done by choosing appropriate $y_{ij}^l$'s such that 
$ \sum_l y_{ij}^l =  \frac{ B_i \SB_j}{\sum_j \SB_j} \qquad \text{and} \qquad y_{ij}^l  \leq B_{ij}^l. $
\end{proof}

Then, following the same steps as those in the proof of Lemma~\ref{uniqueness}, we also show that
the spending vector for spending constraint utilities is unique as well.
\begin{lemma}
For spending constraint utility functions the spending vector $q$ is unique.
\end{lemma}

\subsection{Rationality of SR equilibria under spending constraint utility}
\label{sec:rationalitym1sc}

\begin{lemma} 
	In spending restricted market model under spending constraint utility functions, a rational equilibrium exists 
	if $\sum_j \SB_j \geq \sum_i B_i$  and all the parameters specified are rational numbers. 
	\end{lemma}
	\begin{proof}
		For a buyer $i$ and good $j$, let $S_{ij}^{+}$ be the set of segments $l$ such that $b_{ij}^l = B_{ij}^l$, $S_{ij}^{0}$ be the set of segments such that $B_{ij}^l > b_{ij}^l > 0$, and $S_{ij}^{-}$ be the set of segments such that $b_{ij}^l = 0$. Also, let $\mathcal{S}$ be the family of all $S_{ij}^{+}, S_{ij}^{0}, S_{ij}^{-}$ sets, and $L$ be the set of sellers reaching their earning limits. An equilibrium price $p$, the corresponding spending $b$ and inverse MBB value $\alpha$, if existed, must be a point inside the polyhedron $P(\mathcal{S},L)$ bounded by the following constraints:
		\begin{align*} 
		\forall i \in N, \forall j \in M, \forall l \in S_{ij}^{+} \qquad &  v_{ij}^l \alpha_i \geq  p_j \qquad b_{ij}^l = B_{ij}^l \\
		\forall i \in N, \forall j \in M, \forall l \in S_{ij}^{0} \qquad &  v_{ij}^l \alpha_i =  p_j \qquad  0 \leq b_{ij}^l \leq B_{ij}^l \\
		\forall i \in N, \forall j \in M, \forall l \in S_{ij}^{-} \qquad &  v_{ij}^l \alpha_i \leq  p_j \qquad  b_{ij}^l =0  \\
		\forall i \in N \qquad & \sum_{j,l} b_{ij}^l = B_i \\
		\forall j \in L \qquad & \sum_{i,l} b_{ij}^l= \SB_j \qquad p_j \geq \SB_j \\ 
		\forall j \not \in L \qquad & \sum_{i,l} b_{ij}^l = p_j \qquad p_j \leq \SB_j 
		\end{align*}
		Suppose that all the parameters specified are rational numbers. Again, we can see that a rational equilibrium must also exist if an equilibrium exists. It then follows that a rational equilibrium exists if and only if $\sum_j \SB_j \geq \sum_i B_i$.
		\end{proof}

\section{Utility restricted market model} \label{utility_bound_section}

\subsection{Linear utilities}
The convex program for the linear utility with buyers having utility limits is a natural extension of the Eisenberg-Gale program: 
\begin{equation}\label{cp1} 
\textstyle
\max \sum_{i} B_i \log u_i ~ \text{s.t.} \tag{P3} \\
\end{equation}
\begin{equation}\label{con2.1}
\textstyle
\forall i , \sum_{j} x_{ij} v_{ij} = u_i ,
\end{equation}
\begin{equation} \label{con2.2}
\textstyle
\forall i  , u_i \leq \UB_i ,
\end{equation}
\begin{equation}\label{con2.3}
\textstyle
\forall j  ,\sum_i x_{ij} \leq 1,
\end{equation}
\begin{equation}\label{con2.4}
\textstyle
\forall i,j, x_{ij} \geq 0. 
\end{equation}
In this program, $x_{ij}$ is the amount of good $j$ allocated to buyer $i$, and $u_i$ is the amount of utility that buyer $i$ obtains. Constraint \ref{con2.2} guarantees that the amount of utility buyer $i$ gets does not exceed his utility limit $\UB_i$. 

\begin{lemma}
	\label{lem:cpm2lin} 
Convex program \ref{cp1} captures the equilibrium prices of utility restricted market model under linear utility function.
\end{lemma}

\begin{proof}
	Let $\lambda_i, \mu_i, p_j$ be the dual variables for contraints \ref{con2.1}, \ref{con2.2}, \ref{con2.3} respectively. By the KKT conditions, optimal solutions must satisfy the following:
	\begin{enumerate}
		\item $\forall i \in N,j \in M: \quad -\lambda_i v_{ij}  - p_j \leq 0$
		\item $\forall i \in N,j \in M: \quad x_{ij} >0 \Rightarrow -\lambda_i v_{ij}  - p_j  = 0$
		\item $\forall i \in N : \quad \frac{B_i}{u_i} + \lambda_i - \mu_i= 0$
		\item $\forall i \in N : \quad  \mu_i \geq 0$
		\item $\forall i \in N : \quad  \mu_i > 0 \Rightarrow u_i =  \UB_i $
		\item $\forall j \in N : \quad  p_j \geq 0$
		\item $\forall j \in N : \quad  p_j > 0 \Rightarrow \sum_i x_{ij} =  1 $
	\end{enumerate}
	
	From the first 3 conditions, we have $\forall i \in N,j \in M$ : 
	$ \frac{v_{ij}}{p_j} \leq \frac{u_i}{B_i  - \mu_i u_i}$
	and if $x_{ij} > 0$ then
	$\frac{v_{ij}}{p_j} = \frac{u_i}{B_i - \mu_i u_i}.$
	
	We will show that $p$ is an equilibrium price with allocation $x$. From the above observation, it is easy to see that each buyer $i$ only spends money on his MBB goods at price $p$.
	Moreover, we know that if $p_j >0$ then good $j$ must be fully sold. Therefore, the only remaining thing to prove is that at price $p$ each buyer either spends all his money or attains his utility limit. If $u_i = \UB_i$ then buyer $i$ reaches his utility limit and the amount of money he spends is $B_i - \mu_i \UB_i$, which is at most $B_i$. If $u_i < \UB_i$ then $\mu_i = 0$ and the amount of money he spends is $B_i$.
\end{proof}

We now extend these results to Leontief and CES utility functions.
Utility function $f_i$ is said to be {\em Leontief} if,
given parameters $a_{ij} \in \Rplus \cup \{0\}$ for each good $j \in M$,  
 $f_i(x) = \min_{j \in M} {x_{ij}/a_{ij}}$.
Finally, $f_i$ is said to be {\em constant elasticity of substitution (CES) with parameter $\rho$} if 
given parameters $\alpha_j$ for each good $j \in M$,  
\[f_i(x) = \left( \sum_{j=1}^n \alpha_j x_j^{\rho} \right)^{{\frac{1}{\rho}}}  . \]

\subsection{Utility restricted market model under Leontief utilities}
\label{sec:m2leontief}
The convex program for the Leontief utility model is as follows:
\begin{equation}\label{cp2}
\max  \sum_{i} B_i \log u_i ~ \text{s.t.} \tag{P4} 
\end{equation}
\begin{equation}\label{con3.1}
\forall i,j, u_i \phi_{ij} = x_{ij},
\end{equation}
\begin{equation}\label{con3.2} 
\forall i, u_i \leq \UB_i ,
\end{equation}
\begin{equation} \label{con3.3}
\forall j, \sum_i x_{ij} \leq 1\\
\end{equation}
\begin{equation} \label{con3.4}
\forall i,j,x_{ij} \geq 0.
\end{equation}
\begin{lemma} 
	\label{lem:cpm2leontief}
	Convex program \ref{cp2} captures the equilibrium prices of utility restricted market model under Leontief utility function.
\end{lemma}

\begin{proof}
	Let $\lambda_{ij}, \mu_i, p_j$ be the dual variables for constraints \ref{con3.1}, \ref{con3.2}, \ref{con3.3} respectively. By the KKT conditions, optimal solutions must satisfy the following:
	\begin{enumerate}
		\item $\forall i \in N,j \in M: \quad -\lambda_{ij}  - p_j \leq 0$
		\item $\forall i \in N,j \in M: \quad x_{ij} >0 \Rightarrow -\lambda_{ij}  - p_j = 0$
		\item $\forall i \in N : \quad \frac{B_i}{u_i} + \sum_j \lambda_{ij} \phi_{ij} - \mu_i= 0$
		\item $\forall i \in N : \quad  \mu_i \geq 0$
		\item $\forall i \in N : \quad  \mu_i > 0 \Rightarrow u_i =  \UB_i $
		\item $\forall j \in M : \quad  p_j \geq 0$
		\item $\forall j \in M : \quad  p_j > 0 \Rightarrow \sum_i x_{ij} =  1 $
	\end{enumerate}
	
	Notice that in this model, we may assume that $u_i > 0$ for all $i \in N$. It follows from constraint $\ref{con3.1}$ that $x_{ij} = 0$ if and only if $\phi_{ij} = 0$. From the second KKT condition, we know that if $\phi_{ij} >0$, we must have $ \lambda_{ij} = -p_j$. Substituting in the third condition we have:
	\[  \frac{B_i}{u_i}  - \mu_i = \sum_j p_j \phi_{ij} \]
	Therefore, 
	\[ B_i - \mu_i u_i = \sum_j p_j \phi_{ij} \frac{x_{ij}}{\phi_{ij}} = \sum_j p_j x_{ij}\]
	It follows that $B_i - \mu_i u_i$ is actually the amount of money that buyer $i$ spends. By complementary slackness condition, if $u_i  < \UB_i$ then $\mu_i = 0$ and $i$ spends all his budget. Otherwise, if $u_i = \UB_i$ then $B_i - \mu_i u_i \leq B_i$. Therefore, in this model, a buyer $i$ either spends all his budget or attains his utility limit. Moreover, we know that if $p_j >0$ then good $j$ is fully sold. Thus, $p$ is an equilibrium price with allocation $x$.
\end{proof}

\subsection{Utility restricted marked model under CES utilities}
\label{sec:m2ces}
The convex program for the CES utility model with parameter $\rho$ is as follows: 
\begin{equation} \label{cp3}
\max \sum_{i} B_i \log u_i ~\text{s.t.} \tag{P5}
\end{equation}
\begin{equation} \label{con4.1}
\forall i, u_i = \left( \sum v_{ij} x_{ij}^{\rho} \right)^{\frac{1}{\rho}},
\end{equation}
\begin{equation} \label{con4.2}
\forall i ,  u_i \leq \UB_i,
\end{equation}
\begin{equation} \label{con4.3}
\forall j, \sum_i x_{ij} \leq 1,
\end{equation}
\begin{equation} \label{con4.4}
\forall i,j, x_{ij} \geq 0.
\end{equation}

Notice that in this model, $\partial u_i / \partial x_{ij} = u_i^{1-\rho}v_{ij} x_{ij}^{\rho -1}$ has the same term $u_i^{1-\rho}v_{ij}$ for all $x_{ij}$'s. Moreover, $\partial u_i / \partial x_{ij}$ decreases when  $x_{ij}$ increases. It follows that the best spending strategy for a buyer $i$ is to start with $x_{ij} = 0 \quad \forall j \in M$ and spend money on goods $j$ that maximize the ratio $\frac{\partial u_i / \partial x_{ij} }{p_j}$ at every point. At the end of the procedure, all goods $j$ such that $x_{ij} >0$ will have the same value for $\frac{\partial u_i / \partial x_{ij} }{p_j}$, and that value is the maximum over all goods.

\begin{lemma} 
	\label{lem:cpm2ces}
	Convex program \ref{cp3} captures the equilibrium prices of utility restricted market model under CES utility function.
\end{lemma}

\begin{proof}
	Let $\lambda_{i}, \mu_i, p_j$ be the dual variables for constraints \ref{con4.1}, \ref{con4.2}, \ref{con4.3} respectively. By the KKT conditions, optimal solutions must satisfy the following:
	\begin{enumerate}
		\item $\forall i \in N,j \in M: \quad -\lambda_{i} u_i^{1-\rho}v_{ij} x_{ij}^{\rho -1}  - p_j \leq 0$
		\item $\forall i \in N,j \in M: \quad x_{ij} >0 \Rightarrow -\lambda_{i} u_i^{1-\rho}v_{ij} x_{ij}^{\rho -1}  - p_j = 0$
		\item $\forall i \in N : \quad \frac{B_i}{u_i} + \lambda_{i}  - \mu_i= 0$
		\item $\forall i \in N : \quad  \mu_i \geq 0$
		\item $\forall i \in N : \quad  \mu_i > 0 \Rightarrow u_i =  \UB_i $
		\item $\forall j \in M : \quad  p_j \geq 0$
		\item $\forall j \in M : \quad  p_j > 0 \Rightarrow \sum_i x_{ij} =  1 $
	\end{enumerate}
	
	We will prove that $p$ is an equilibrium price with allocation $x$.  From the first there KKT conditions, we have
	\[ \frac{u_i^{1-\rho}v_{ij} x_{ij}^{\rho -1}}{p_j} \leq \frac{u_i}{B_i - \mu_i u_i} \]
	and equality happens when $x_{ij} > 0$. Therefore, $x$ is in agreement with the best spending strategy of the buyers, which says that for each buyer $i$, if $x_{ij} >0$ then  $\frac{\partial u_i / \partial x_{ij} }{p_j}$ is maximized over all $j$'s. Moreover, we can see that $B_i - \mu_i u_i$ is the amount of money buyer $i$ spends.  By complementary slackness condition, if $u_i  < \UB_i$ then $\mu_i = 0$ and $i$ spends all his budget. Otherwise, if $u_i = \UB_i$ then $B_i - \mu_i u_i \leq B_i$. Therefore, in this model, a buyer $i$ either spends all his budget or attains his utility limit. Moreover, we know that if $p_j >0$ then good $j$ is fully sold. Thus, $p$ is an equilibrium price with allocation $x$.
\end{proof}

\subsection{Rationality of equilibria for UR market model under linear utilities}
\label{sec:rationalitym2sc}

\begin{lemma} \label{utility_rational}
In UR market model under linear utility functions, a rational equilibrium exists if all the parameters specified are 
rational numbers.
\end{lemma}

\begin{proof}
 Let $A_i$ be the set of goods that buyer $i$ spends money on, $\mathcal{A}$ be the family of $A_i$'s, and $L$ be the set of buyers reaching their utility limits. An equilibrium price $p$, the corresponding spending $b$ and inverse MBB value $\alpha$, if existed, must be a point inside the polyhedron $P(\mathcal{A},L)$ bounded by the following constraints:
\begin{align*} 
\textstyle 
\forall i \in N, \forall j \in A_i \qquad &  v_{ij} \alpha_i =  p_j \\
\textstyle 
\forall j \in M \qquad & v_{ij} \alpha_i \leq p_j \\
\textstyle 
\forall i \in N, \forall j \not \in A_i \qquad & b_{ij} = 0  \\
\textstyle 
\forall j \in N \qquad & \sum_i b_{ij} = p_j \\
\textstyle 
\forall i \in L \qquad & \sum_{j} b_{ij} = \alpha_i \UB_i \qquad \sum_j b_{ij} \leq B_i \\ 
\textstyle 
\forall i \not \in L \qquad & \sum_{j} q_{ij} \leq \alpha_i \UB_i \qquad \sum_j b_{ij} = B_i  \\ 
\textstyle 
\forall i \in N, j \in M \qquad & b_{ij} \geq 0
\end{align*}
Suppose that all the parameters specified in this model are rational numbers. By a similar argument to Lemma \ref{earning_rational}, we can see that an equilibrium exists if and only if a rational equilibrium exists. It follow from Lemma \ref{utility_existence} that a rational equilibrium price must always exist if all the parameters specified are rational numbers.
\end{proof}

\subsection{Existence and Uniqueness of UR equilibrium}
For UR market model, we show that an equilibrium always exists for all utility functions we mentioned in the previous section. On the uniqueness side, the utility vector is unique.
To verify that the price vector is not unique, consider a market with only one buyer with utility function $u(x)=x_1$ and one seller. Let $\UB_1=1$ and $B_1=2$. 
It is easy to see every price in interval $[1,2]$ is an equilibrium price.  

\begin{lemma} \label{utility_existence}
In UR market model under linear, Leontief and CES utility functions, an equilibrium price always exists. 
\end{lemma}
\begin{proof} An equilibrium price exists if and only if the feasible region of the convex program is not empty. In \ref{cp1}, \ref{cp2} and \ref{cp3}, $x_{ij} = 0$ for all $i,j$ is a feasible solution. Therefore, the feasible region is not empty and an equilibrium exists.
\end{proof}
\begin{lemma}
In UR market model under linear, Leontief and CES utility functions, the utilities of an equilibrium are unique.
\end{lemma}
\begin{proof}
In section \ref{utility_bound_section}, we showed every equilibrium correspond to an solution of a convex program with an objective function of the form $\sum_{i} B_i \log u_i$.
It is easy to see that the objective function is strictly concave.
Therefore, there is a unique vector $u$ that maximizes the objective function and the lemma follows.
\end{proof}
\section{Proofs of Theorem~\ref{thm:approx} and Lemma~\ref{thm:approx_lb} (Approximation Factor Bounds)}\label{app:approx}

\subsection{Approximation Factor Upper Bound}
For each item $j$ that has more than one child-agent in the spending 
graph $Q(b)$, remove the edges connecting it to all but the one child-agent 
that spends the most money on $j$, i.e., the one with the largest $b_{ij}$ value. 
This yields a pruned spending graph $P(b)$ that is also a forest of trees.
We refer to the trees of the pruned graph $P(b)$ as the {\em matching-trees}. 
In every matching-tree $T$ with $k\geq 2$ agents, when the algorithm
reaches its last step, every remaining item has exactly one parent-agent 
and one child-agent, so all but one agent can be matched to one of these
items. Our proof shows that there exists a matching of the remaining items 
such that the agents within $T$ have a ``high'' NSW.

A naive way to match the agents in the last step of the algorithm would be 
to match all of them, except the one that has accrued the highest value
during the previous steps. It was already observed in~\cite{CG15} that, 
for any matching-tree $T$ of $k$ agents, there exists an agent who was 
assigned value at least $1/(2k)$ during Steps 3 and 4 of the algorithm,
so we could match every agent in $T$, except him. But, what is the worst
case distribution of value that can arise in this matching?

If $T$ is some matching-tree of the pruned spending graph $P(b)$, then 
let $M_T$ denote the union of items in $T$ with the items that were assigned 
to agents in $T$ in Steps 3 and 4. Also, let $H$ be the set of items with
$p_j>1$ in the SR equilibrium and $H_T$ the subset of those items that belong
to $T$. In proving this theorem, we use the following lemma from~\cite{CG15}.

\begin{lemma}[\cite{CG15}]\label{lem:Tproperties}
For any matching-tree $T$ with $k$ agents, there exists an agent $i\in T$ 
who, during Steps 3 and 4 received one or more items that she values at least $1/(2k)$. 
Also, for items in $M_T$:
\begin{equation}\label{ineq:total_value}
\sum_{j\in M_T}q_j \geq k-\frac{1}{2}.
\end{equation}
\end{lemma}

Let $x'$ be the integral allocation that would arise if we follow the SRR algorithm
up to Step 4, and then use the naive matching described above. For simplicity, we
assume that the valuations of the agents are scaled in such a way that $v_{ij}= p_j$ 
if $b_{ij}>0$, which allows us to use {\srub} as an upper bound of OPT.
We begin by showing that, if every agent receives a value of at least $1/2$ in
$x'$, then the theorem follows. To verify this fact, note that every agent who is
matched to an item $j$ with price $p_j>1$ has a value of at least $p_j$, and every
other agent has a value of at least $1/2$, so
\[\left(\prod_i v_i(x')\right)^{1/n} ~\geq~ \left(\frac{1}{2^n} \cdot \prod_{j\in H} p_j \right)^{1/n} ~\geq~ \frac{1}{2} \cdot \left(\prod_i v_i(x^*)\right)^{1/n}.\]

For any matching-tree with $k=1$ agent, Inequality~\eqref{ineq:total_value} implies that
this agent will receive value at least $k-1/2=1/2$. Therefore, we now, assume that 
there exists some matching-tree $T$ with $k\geq 2$ agents such that
some agent $\alpha$ in $T$ gets a value less than $1/2$ in $x'$. Let $v_{\alpha}(x')$, 
or $v_{\alpha}$ for short, be the value that this agent receives. Since $v_{\alpha}<1/2$, 
this agent is the only one in $T$ that was not matched to an item with $p_j>1/2$, so
every other agent $i$ in $T$ has $v_i(x')\geq 1/2$.

\begin{lemma}\label{lem:heavy_agents}
Over all the possible allocations $x'$, the one with the minimum product of the
valuations, has at least $\left\lfloor\frac{k-2v_{\alpha}}{1+2v_{\alpha}}\right\rfloor$ 
agents with value $v_i(x')\geq 1$.
\end{lemma}
\begin{proof}
Let $k_1$ be the number of agents with value at least 1 in $x'$, and assume that 
$k_1\leq \left\lfloor\frac{k-2v_{\alpha}}{1+2v_{\alpha}}\right\rfloor -1$. 
Since every agent other than $\alpha$ was matched to an item with $q_j\geq 1/2$,
we know that the value of the $k_1$ agents before the matching was at most 
$v_{\alpha}$. Hence, for each such agent $i$ the sum of the $q_j$ values of the 
items that were assigned to $i$ in $x'$ is at most $1+v_{\alpha}$. As a result, 
if $M'$ is the union of items that were assigned to agent $\alpha$ and the $k_1$ agents,
we know that $\sum_{j\in M'}q_j\leq k_1(1+v_{\alpha})+v_{\alpha}$.

Using Inequality~\eqref{ineq:total_value}, we get
\[\sum_{j\in M\setminus M'} q_j ~\geq~ k-\frac{1}{2}-(k_1(1+v_{\alpha})+v_{\alpha}) 
~=~ \frac{1}{2}(k-k_1) + \frac{1}{2}k -\left(\frac{1}{2}+v_{\alpha}\right)(k_1+1).\]
But, we have assumed that $k_1+1\leq \left\lfloor\frac{k-2v_{\alpha}}{1+2v_{\alpha}}\right\rfloor 
\leq \frac{k-2v_{\alpha}}{1+2v_{\alpha}}$, so
\begin{equation*}
\sum_{j\in M\setminus M'} q_j ~\geq~ \frac{1}{2}(k-k_1) + \frac{1}{2}k -\frac{k-2v_{\alpha}}{2} 
~=~ \frac{1}{2}(k-k_1) + v_{\alpha}.
\end{equation*}
Therefore, the remaining $k-k_1-1$ agents have a total value more than $(k-k_1)/2$,
i.e., strictly more than 1/2 on average. It also implies that at least two of these 
agents have value strictly more than 1/2. If $k-k_1-2$ agents had value equal to 
1/2 then the remaining agent would have a value more than $\frac{1}{2}(k-k_1) - 
\frac{1}{2}(k-k_1-2)=1$, which contradicts our assumption that only $k_1$ agents 
have value at least 1. Let $v_1,v_2\in (1/2, 1)$ be the values of two such agents
in the worst case outcome. It is then easy to verify that, if we were to instead
give value $1/2$ to the one agent and $v_1+v_2-1/2$ to the other, the NSW would
drop. This contradicts our assumption that this is a worst case outcome.
\end{proof}

\begin{lemma}\label{ineq:bound1}
For any matching-tree $T$ with $k$ agents, the allocation $x$ of the SRR algorithm satisfies
\[\prod_{i\in T} v_i(x') ~\geq~ \frac{1}{2^k} \prod_{j\in H_T} p_j.\]
\end{lemma}
\begin{proof}
%
Let $k_1$ be the number of agents with $v_i(x')\geq 1$. Given any agent $i$ among these 
$k_1$ agents, if $j$ is the item that he was matched to, then has value $v_i(x')\geq \max\{1, p_j\}$. 
As a result, the product of the values of these $k_1$ players is at least $\prod_{j\in H_T} p_j$.
Therefore, it suffices to show that the product of the remaining $k-k_1$ agents is at least 
$1/2^k$.

Among the $k_2 = k-k_1-1$ agents that get value in $[1/2, 1)$, it is easy to verify that
their product is minimized when at most one agent among them gets value higher than $1/2$.
If we let $v_{\beta}$ be the value of that player, and using Inequality~\eqref{ineq:total_value}, 
we get
\begin{equation}\label{ineq:vbeta_bound}
v_{\beta} ~\geq~ k-\frac{1}{2}-\left[ k_1 (1+v_{\alpha})+ v_{\alpha}+ (k-k_1-2)\frac{1}{2} \right].
\end{equation}
If we let $\overline{k}_1 =\frac{k-2v_{\alpha}}{1+2v_{\alpha}}$ and $\delta= \overline{k}_1- k_1
= \frac{k-2v_{\alpha}}{1+2v_{\alpha}} -\left\lfloor\frac{k-2v_{\alpha}}{1+2v_{\alpha}}\right\rfloor$ be
the rounding error, then Inequality~\eqref{ineq:vbeta_bound} yields
\[v_{\beta} ~\geq~ \frac{\delta+1}{2}.\]
This implies that
\[\prod_{i\in T} v_i(x') ~\geq~ \frac{v_{\alpha} v_{\beta}}{2^{k-k_1-2}} \prod_{j\in H_T} p_j
~\geq~ \frac{v_{\alpha}}{2^{k-\overline{k}_1-1}} \frac{\delta+1}{2^{\delta}} \prod_{j\in H_T} p_j
~\geq~ \frac{v_{\alpha}}{2^{k-\overline{k}_1-1}} \prod_{j\in H_T} p_j,\]
where the last inequality comes from the fact that $\delta+1\geq 2^{\delta}$
for $\delta\in [0,1]$. To verify this fact note that $(\delta+1 - 2^{\delta})''=-2^{\delta}\ln^2 2<0$,
so this is minimized at either $\delta=0$ or $\delta=1$, both of which yield $\delta+1 - 2^{\delta}= 0$.
To conclude the proof, it suffices to show that for every $v_{\alpha}\in [1/(2k), 1/2]$ and every
$k\geq 2$ we have 
\[\frac{v_{\alpha}}{2^{k-\overline{k}_1-1}}\geq \frac{1}{2^k} \quad\text{or, equivalently,}\quad v_{\alpha} 2^{\frac{k+1}{1+2v_{\alpha}}}\geq 1.\]
For $k\geq 7$, it is easy to verify that this inequality holds. In particular, using the fact that $v_{\alpha}\in [1/(2k), 1/2]$,
\[v_{\alpha} 2^{\frac{k+1}{1+2v_{\alpha}}} ~\geq~ \frac{1}{2k} 2^{\frac{k+1}{1+2\frac{1}{2}}} ~\geq~ \frac{2^{\frac{k+1}{2}}}{2k} ~\geq 1,  \quad\text{for } k\geq 7.\]

Note that $v_{\alpha} 2^{\frac{k+1}{1+2v_{\alpha}}}$ is minimized at the same points as $\log v_{\alpha} + \frac{k+1}{1+2v_{\alpha}}$. Taking a derivative
w.r.t.\ $v_{\alpha}$ gives
\[\left(\log v_{\alpha} + \frac{k+1}{1+2v_{\alpha}}\right)' 
~=~ \frac{1}{ v_{\alpha} \ln 2} -  \frac{2(k+1)}{(1+2v_{\alpha})^2}.\]
For $k\leq 4$, this derivative is positive for any value of $v_{\alpha}$, so
$v_{\alpha} 2^{\frac{k+1}{1+2v_{\alpha}}}$ is minimized at $v_{\alpha}=1/(2k)$, 
where its value is equal to $\frac{1}{2k} 2^{\frac{k(k+1)}{k+1}} = \frac{2^k}{2k}\geq 1$.
Finally, replacing $k=5$ and $k=6$ and minimizing $v_{\alpha} 2^{\frac{k+1}{1+2v_{\alpha}}}$ over all 
values of $v_{\alpha}$ also shows that this function is minimized at $v_{\alpha}=1/10$ and $v_{\alpha}=1/12$
respectively, and its value is at least $1$, which concludes the proof.
\end{proof}

The inequality of Lemma~\ref{ineq:bound1} implies the desired approximation factor if we observe that
\[\left(\prod_i v_i(x)\right)^{1/n} ~=~ \left( \prod_T \prod_{i\in T} v_i(x)\right)^{1/n} ~\geq~ \frac{1}{2} \left(\prod_{j\in H} p_j \right)^{1/n}.\]

\subsection{Approximation Factor Lower Bound}
\begin{proof}
Consider an instance with $m=2\kappa$ items and $n=\kappa +1$ agents.
Each agent $i\in [1,\kappa]$ has a value of $1/2$ for item $i$ and
a value of $1/2+1/\kappa$ for item $2i$. The value of these agents for 
every other item is $0$. Finally, agent $\kappa+1$ values
every item from $1$ to $\kappa$ for a value of $1$ and has value $0$ for the
rest. The item prices in the SR equilibrium for this instance are $1/2$
for the first $\kappa$ items and $1/2+1/\kappa$ for the remaining $\kappa$
items. Agent $\kappa+1$ spends $1/\kappa$ on each one of the first $\kappa$
items, while each agent $i\in [1,\kappa]$ spends $1-1/\kappa$ on item $i$
and his remaining budget of $1+1/\kappa$ on item $2i$. 

Facing this SR equilibrium, assume that the SRR algorithm chooses agent $\kappa+1$
as the root-agent in Step 2, then it would assign all of the first $\kappa$ 
items to this agent. To verify this fact note that for every item $j$ among
the first $\kappa$ ones, $q_j =1/2$ and agent $\kappa+1$ is the parent-agent.
On the other hand, every other agent $i\in [1, \kappa]$ would get only 
item $2i$. This leads to a product of valuations equal to 
$\frac{\kappa}{\left(2+1/\kappa\right)^\kappa}$. If, on the other hand, agent 
$\kappa+1$ was allocated just one of the first $\kappa$ items and gave each of 
the other $\kappa-1$ items to the agents that value them, the product of the 
valuations would be more than $\frac{1}{2}$. For large values of $\kappa$ the 
ratio between the NSW of these two outcomes converges to 2. Finally, note
that, even if the algorithm chose some different agent as the root, the result
would not be affected in the limit.
%
%
%
\end{proof}

\end{document}